\documentclass[aps,nofootinbib,amsmath,prd,onecolumn,notitlepage,showpacs,superscriptaddress,11pt]{revtex4-1}

\usepackage[american]{babel}
\usepackage{amsfonts}
\usepackage{amsmath}
\usepackage{amssymb}
\usepackage{slashed}
\usepackage{xcolor}
\usepackage{bm}
\usepackage{float}
\usepackage[utf8]{inputenc}
\usepackage{nicefrac}
\usepackage{hyperref}
\usepackage{url}
\usepackage{tikz-cd}
\usepackage[normalem]{ulem}

\definecolor{darkblue}{rgb}{0.1,0.1,0.7}
\hypersetup{colorlinks,
           linkcolor={darkblue},
           citecolor={darkblue},
           urlcolor={darkblue}
}

\newcommand{\lvec}[2]{\raise #1\hbox{$^\leftarrow$} \hspace{-9pt} #2}
\newcommand{\rvec}[2]{\raise #1\hbox{$^\rightarrow$} \hspace{-9pt} #2}
\newcommand{\lrvec}[2]{\raise #1\hbox{$^\leftrightarrow$} \hspace{-9pt} #2}

\allowdisplaybreaks

\usepackage[normalem]{ulem}

\allowdisplaybreaks

\begin{document}

\title{Substructures of the Weyl group and their physical applications}

\author{Riccardo Martini}
\email{riccardo.martini@bo.infn.it}
\affiliation{Dipartimento di Fisica e Astronomia, Universit\`a di Bologna, via Irnerio 46, 40126 Bologna, Italy}
\affiliation{INFN - Sezione di Bologna, via Irnerio 46, 40126 Bologna, Italy}

\author{Gregorio Paci}
\email{gregorio.paci@phd.unipi.it}
\affiliation{
Dipartimento di Fisica, Universit\`a di Pisa, Largo Bruno Pontecorvo 3, 56127 Pisa, Italy}
\affiliation{INFN - Sezione di Pisa, Largo Bruno Pontecorvo 3, 56127 Pisa, Italy}

\author{\\Dario Sauro}
\email{dario.sauro@phd.unipi.it}
\affiliation{
Dipartimento di Fisica, Universit\`a di Pisa, Largo Bruno Pontecorvo 3, 56127 Pisa, Italy}
\affiliation{INFN - Sezione di Pisa, Largo Bruno Pontecorvo 3, 56127 Pisa, Italy}

\author{Gian Paolo Vacca}
\email{vacca@bo.infn.it}
\affiliation{INFN - Sezione di Bologna, via Irnerio 46, 40126 Bologna, Italy}

\author{Omar Zanusso}
\email{omar.zanusso@unipi.it}
\affiliation{
Dipartimento di Fisica, Universit\`a di Pisa, Largo Bruno Pontecorvo 3, 56127 Pisa, Italy}
\affiliation{INFN - Sezione di Pisa, Largo Bruno Pontecorvo 3, 56127 Pisa, Italy}

\begin{abstract}
%
We study substructures of the Weyl group of conformal transformations of the metric of (pseudo)Riemannian manifolds. These substructures are identified by differential constraints on the conformal factors of the transformations
which are chosen such that their composition is associative.
Mathematically, apart from rare exceptions, they are partial associative groupoids, not groups, so they do not have an algebra of infinitesimal transformations, but this limitation can be partially circumvented using some of their properties cleverly.
We classify and discuss the substructures with two-derivatives differential constraints, the most famous of which being known as the harmonic or restricted Weyl group in the physics literature,
but we also show the existence of a lightcone constraint which realizes a proper subgroup of the Weyl group.
We then show the physical implications that come from invariance under the two most important substructures, concentrating on classical properties of the energy-momentum tensor and a generalization of the quantum trace anomaly.
We also elaborate further on the harmonic substructure, which can be interpreted as partial gauge fixing of full Weyl invariance using BRST methods.
Finally, we discuss how to construct differential constraints of arbitrary higher-derivative order and present, as examples, generalizations involving scalar constraints with four and six derivatives.
\end{abstract}

\pacs{}
\maketitle

\section{The Weyl group and its substructures}\label{sect:intro}

Consider a (pseudo)Riemannian manifold equipped with some metric $g_{\mu\nu}$ and the covariant density abbreviated as
$\sqrt{g} \equiv (|\det(g_{\mu\nu})|)^{1/2}$.
In this paper we refer to the (full) \emph{Weyl group} as the group of transformations that acts
on any metric as
\begin{equation}\label{eq:weyl}
 g_{\mu\nu} \to g'_{\mu\nu}= \Omega^2 g_{\mu\nu}\,,
\end{equation}
for some local function $\Omega=\Omega(x)$ different than zero.
The Weyl group is obviously Abelian and the inverse transformation of \eqref{eq:weyl} is guaranteed to exist as
$g_{\mu\nu} \to g'_{\mu\nu}=\Omega^{-2}g_{\mu\nu}$.
Infinitesimal
transformations can be obtained by setting $\Omega=1+\sigma$, for some infinitesimal function $\sigma=\sigma(x)\ll 1$, which also provide the algebra of the continuous group. The algebra of infinitesimal
transformation is then generated by $\delta_\sigma$, where $\delta_\sigma g_{\mu\nu}=2\sigma g_{\mu\nu}$
and the Abelian nature of the group is reflected in the commutativity of the algebra itself, i.e., $[\delta_\sigma,\delta_{\sigma'}]=0$.

As for the well-known physical implications of Weyl symmetry, a classical action $S[\Psi]$ of some field $\Psi$ that transforms with a given weight $w_\Psi$ under the Weyl group, i.e., $\Psi \to \Psi'=\Omega^{w_\Psi}\Psi$ and infinitesimally $\delta_\sigma\Psi = w_\Psi \sigma \Psi$,
has the property that the variational energy-momentum tensor (EMT) $T_{\mu\nu}$
is traceless on-shell
\begin{equation}\label{eq:emt-weyl}
 T^{\mu\nu} \equiv -\frac{2}{\sqrt{g}}\frac{\delta S[\Psi]}{\delta g_{\mu\nu}}\,,
 \qquad\qquad
 T^\mu{}_\mu|_{\rm on-shell} =0\,,
\end{equation}
because of the N\"other identities.
Here on-shell means that, in order to prove the tracelessness of the EMT, we must make use of the equations of motion of the field $\Psi$, that is, we must enforce $\frac{\delta S}{\delta \Psi}=0$, which is a well-known fact. Quantum mechanically the Weyl symmetry may be anomalous, in which case $\langle T^\mu{}_\mu\rangle \neq 0$ even on-shell, where
\begin{equation}\label{eq:emt-weyl-quantum}
 \langle T^{\mu\nu} \rangle \equiv -\frac{2}{\sqrt{g}}\frac{\delta \Gamma[\Psi]}{\delta g_{\mu\nu}}\,,
 \qquad\qquad
 \langle T^\mu{}_\mu\rangle|_{\rm on-shell} = {\cal A} + \bigl(\beta ~ {\rm terms}\bigr)\,,
\end{equation}
given $\Gamma[\Psi]$, which is the renormalized action of the renormalized field $\Psi$, here denoted with the same symbol as the classical field for convenience,
and ${\cal A}$, which is the conformal anomaly \cite{Deser:1993yx} satisfying the appropriate Wess-Zumino \cite{Wess:1971yu} integrability conditions \cite{Bonora:1983ff}. The ``$\beta$ terms'' schematically represent the scale dependence induced by eventual renormalized running couplings through their beta functions and is generally zero at renormalization group fixed points \cite{Osborn:1991gm}.

In this paper we are concerned with the existence and the physical consequences of some associative substructures of the Weyl group \cite{Iorio:1996ad}, i.e., group-like subsets
for which the function $\Omega(x)$ satisfies some \emph{differential constraints} that depend on the metric $g_{\mu\nu}$ itself.\footnote{%
A concise overview of group-like structures arranged in the form of a table according to their properties is given in Wikipedia~\cite{wikipedia}.
}
An example of differential constraint would be
$$\Box_g\Omega=0\,,$$
which gives an associative substructure of the Weyl group in (and only in) $d=4$,\footnote{This constraint needs to be nonlinearly deformed in arbitrary dimension $d$ for associativity to hold, as shown later in Eq.~\eqref{gen_harmonic}.} and becomes quite relevant in the rest of the paper under the name of harmonic (or restricted) Weyl group \cite{Iorio:1996ad}. Importantly, these group-like structures are required to be closed and associative like the full Weyl group. We concentrate mainly on classical theories, but also indulge on their quantum mechanical counterparts as well, while attempting to give a meaningful generalization to the notion of anomaly.

Given that the covariant differential constraints on $\Omega(x)$ depend on $g_{\mu\nu}$ (e.g., $\Box_g\Omega=0$ depends on $g_{\mu\nu}$ through the covariant Laplacian), such structures are, for the most part, not true subgroups, nor they actually are groups. Rather they can be understood as \emph{partial associative groupoids}, also known as partial semigroupoids \cite{mathoverflow}, as we clarify in the following and in appendix \ref{sect:appendix-groupoids}. 
For simplicity, we often still refer to them simply as ``groupoids'' or even improperly as ``groups'' or ``subgroups''
as was done in the existing physics literature \cite{Iorio:1996ad}, but the distinction is relevant for the applications of this paper and should always be kept in mind.

To our knowledge, two such substructures, that are associated to independent second order derivative constraints, have been investigated before and used in both the physics and math literatures under the names of ``restricted Weyl group'' and ``Liouville transformations'' (e.g., Refs.~\cite{Iorio:1996ad,Oda:2020wdd,Edery:2014nha,Kuhnel-Rademacher}). We briefly review them later in this section, discussing some physical aspects, and giving also preliminar ideas that are useful for generalizations to higher derivative constraints.

The paper is organized as follows: in the rest of Sect.~\ref{sect:intro} we list the two aforementioned interesting substructures, discussing briefly their associative properties, commenting also on the fact that other potential constraints leads to trivial structures. In Sect.~\ref{sect:constructions} we show the uniqueness of some structures as partial associative groupoids defined by differential constraints with two derivatives on the function $\Omega$. 
Our classification also shows the existence of a special subgroup structure associated to a lightcone scalar constraint on derivatives of $\Omega$.
In Sect.~\ref{sect:implications-and-anomaly} we show their physical implications in terms of properties of the energy-momentum tensor, both classically and, using two examples, also quantum mechanically.
In Sect.~\ref{sect:general-recipe} we show how to construct general field theories of arbitrary tensors that are invariant under the subgroups of the Weyl group, generalizing a know method that is often used to construct fully Weyl invariant theories.
In Sect.~\ref{sect:partial-gf} we show how models with a dynamical metric, that are invariant under the restricted Weyl subgroup, can be seen as partial gauge fixings of models with full conformal/Weyl invariance. This is done applying BRST methods and follows, in part, previous constructions that have already appeared in the literature \cite{Oda:2020wdd,Edery:2014nha}.
In Sect.~\ref{sect:generalization} we discuss a general way to systematically construct higher derivative constraints and present the form of the general scalar constraints with four derivatives. Furthermore, we provide a specific example of a constraint with six derivatives motivated by the study of conformal geometry.
In Sect.~\ref{sect:conclusions} we summarize our results and give some conclusions.
Appendix~\ref{sect:appendix-groupoids} is more mathematical in nature and discusses the substructures as partial semigroupoids, while Appendix~\ref{sect:appendix-cohomology} gives a brief cohomological analysis of the anomaly of the harmonic substructure.

\subsection{The harmonic Weyl group}\label{sect:harmonic-group}

The first and arguably simplest substructure
is the \emph{harmonic Weyl group}, which
is also known as the \emph{restricted Weyl group} \cite{Iorio:1996ad,Oda:2020wdd,Kamimura:2021wzf,Oda:2021kvx}, but recall that it actually is a groupoid.
When first reading this section we recommend to concentrate on relations and properties, rather than on the formal details of the mathematical definition of grupoid, because the former are closely connected to the physical implications discussed later, while the latter can be settled at a later stage reading Sect.~\ref{subsec6A} and Appendix~\ref{sect:appendix-groupoids}.
The harmonic Weyl group is defined as the set of transformations
\begin{equation}\label{eq:harmonic}
 g_{\mu\nu} \to g'_{\mu\nu}= \Omega^2 g_{\mu\nu}\,, \qquad {\rm for} \qquad H_g(\Omega)=0\,,
\end{equation}
where the differential constraint $H_g(\Omega)$ depends on the ``starting point'' metric $g_{\mu\nu}$ and is defined
\begin{equation}
\label{gen_harmonic}
H_g(\Omega)=\Box_g \Omega + \frac{d-4}{2\Omega} g^{\mu\nu}\partial_\mu\Omega\partial_\nu\Omega
\end{equation}
with
$\Box_g =g^{\mu\nu}\nabla_\mu\nabla_\nu$ being the covariant Laplacian, and $\nabla_\mu$ being the unique Levi-Civita symmetric connection compatible with $g_{\mu\nu}$. Notice that $\nabla_\mu \Omega = \partial_\mu \Omega$, so we can denote $\Box_g\Omega=\nabla^\mu\partial_\mu \Omega$. We also adopt the notation $\nabla^\mu \Omega=\partial^\mu \Omega=g^{\mu\nu} \partial_\nu \Omega$, so it must be kept in mind that $\partial^\mu$ secretly depends on the inverse metric, which is affected by Weyl transformations. Furthermore, expanding all terms, we have that
$\Box_g\Omega=\partial^\mu\partial_\mu \Omega-g^{\mu\nu}\Gamma_{\mu}{}^{\rho}{}_\nu \partial_\rho \Omega$, where $\Gamma_{\mu}{}^{\rho}{}_\nu=\frac{1}{2}g^{\rho\theta}(\partial_{\mu} g_{\theta\nu}+\partial_{\nu} g_{\mu\theta}-\partial_{\theta} g_{\mu\nu})$ are the components of the unique Levi-Civita connection of $g_{\mu\nu}$.
Importantly, in $d=4$ the harmonic condition \eqref{eq:harmonic} becomes linear in $\Omega$ and simplifies to $H_g(\Omega)=\Box_g\Omega=0$, hence the name ``harmonic''.

As mentioned above, the harmonic Weyl group is not a subgroup of
the Weyl group in the mathematical sense, but, rather, a groupoid of the Weyl group on the space of all metrics, because the condition on the conformal factor in \eqref{eq:harmonic} depends on the metric $g_{\mu\nu}$ itself.
In fact, it is closed under group multiplication, meaning that two consecutive harmonic transformations combine into a harmonic transformation, although this is less trivial than it sounds.
To prove closure it is necessary to show that, given the transformations
\begin{equation}\label{eq:omega1}
 g_{\mu\nu} \to g'_{\mu\nu}=(\Omega_1)^2 g_{\mu\nu} \qquad \qquad H_g(\Omega_1)=0
\end{equation}
and 
\begin{equation}\label{eq:omega2}
 g'_{\mu\nu} \to g''_{\mu\nu}=(\Omega_2)^2 g'_{\mu\nu} \qquad \qquad H_{g'}(\Omega_2)=0\,,
\end{equation}
which act on the metrics $g_{\mu\nu}$ and $g'_{\mu\nu}=(\Omega_1)^2 g_{\mu\nu}$ respectively,
their combination gives a third harmonic transformation such that, denoting $\Omega_3 = \Omega_1\cdot \Omega_2$ the pointwise product of the two functions,
\begin{equation}\label{eq:omega3}
 g_{\mu\nu} \to g''_{\mu\nu}=(\Omega_1 \cdot \Omega_2)^2 g_{\mu\nu} \qquad \qquad H_{g}(\Omega_1 \cdot \Omega_2)=0\,,
\end{equation}
which starts from the first metric $g_{\mu\nu}$. The proof in general $d$ is tedious, but relatively straightforward. We do not report it here to avoid redundancy. In fact, in Sect.~\ref{sect:constructions} we discuss the general construction -- based on the requirement of associativity -- of this type of groupoids in greater detail and the above associativity property follows trivially from the results of Sect.~\ref{sect:constructions}.

The fact that the harmonic Weyl group is not a subgroup of the Weyl group, but rather a group-like structure, should be evident from the definition of the second transformation, which acts \emph{only} on a metric that is rescaled by the action of the first transformation.
Similarly, we can infer the same by inspecting the inverse transformation: if $g'_{\mu\nu}=\Omega^2 g_{\mu\nu}$ with $H_g(\Omega)=0$, we have the inverse transformation $g'_{\mu\nu}=\Omega^{-2}g_{\mu\nu}$ with $H_{\Omega^2 g}(\Omega^{-1})=0$.
One can show in general that, for nonzero $\Omega$, the conditions $H_g(\Omega)=0$
and $H_{\Omega^2 g}(\Omega^{-1})=0$ are completely equivalent for nondegenerate $\Omega$, which guarantees the existence of the inverse transformation. 
On the other hand given $\Omega$ such that $H_g(\Omega)=0$ one has $H_{g}(\Omega^{-1})\ne 0$ implying that the inverse action applied to the same metric is not an element of a group-like structure.
The transformation $g'_{\mu\nu}=\Omega^{-2}g_{\mu\nu}$ with $H_{\Omega^2 g}(\Omega^{-1})=0$ can be either seen as a \emph{right-inverse} transformation (it gives the identity only if applied after the first transformation) or as a \emph{pseudo-inverse}, see also appendix \ref{sect:appendix-groupoids}.

Nevertheless, several properties of groups are shared by groupoids.
In particular, it is trivial to show the existence of the ``identity elements'' $e_g$, given by $\Omega(x)=1$
\begin{equation}\label{eq:identity_element}
 g_{\mu\nu} \to g'_{\mu\nu}= g_{\mu\nu} \qquad \qquad H_{g}(1)=0\,,
\end{equation}
and the label $g$ in $e_g$ tells us on which metric the trivial transformation acts.
Furthermore, the composition in the set of harmonic Weyl transformations, as defined in \eqref{eq:harmonic}, enjoys associativity 
\begin{equation}\label{eq:associativity}
(a \circ b) \circ c = a \circ ( b \circ c ) \,,
\end{equation}
which follows from the composition rule relatively straightforwardly.
For a brief, more mathematical, explanation of the groupoids we again refer to appendix \ref{sect:appendix-groupoids},
where we also discuss the nature of the identity elements, interpreted a set of transformations of the groupoid.

Now recall that the curvature scalar $R$ under general finite Weyl transformations transforms as
\begin{equation}\label{eq:weyl-transformation-R}
 \begin{split}
 R \to R' &= \Omega^{-2}\Bigl(
 R - 2(d-1)\Omega^{-1}\Box_g \Omega-\Omega^{-2} (d-1)(d-4)\partial_\mu \Omega \partial^\mu \Omega
 \Bigr)
 \\
 &= \Omega^{-2}\Bigl(
 R - 2(d-1)\Omega^{-1} H_g (\Omega)
 \Bigr)\,.
 \end{split}
\end{equation}
The important point of \eqref{eq:weyl-transformation-R} is that
the curvature scalar $R$ transforms homogeneously under the harmonic group in general $d\geq 2$, so
\begin{equation}
 R \to R' = \Omega^{-2} R\,, \qquad \qquad {\rm if} ~~ H_g(\Omega)=0\,.
\end{equation}
One immediate consequence is that $R$ can be used to construct an harmonic invariant density $\sqrt{g}R^2$ in $d=4$, where we recall that also the harmonic condition simplifies considerably to $\Box_g \Omega=0$.
The harmonic invariant density thus behaves like
the fully Weyl invariant density $\sqrt{g} W^2$ that is constructed from the square of the Weyl tensor $W_{\mu\nu\alpha\beta}$.
In practice we have that any density
\begin{equation}
 \sqrt{g}\bigl(\alpha_1 R^2 + \alpha_2 W^2 \bigr)
\end{equation}
is harmonic Weyl invariant for arbitrary couplings $\alpha_1$ and $\alpha_2$ in $d=4$.
In Sect.~\ref{sect:partial-gf} we indulge on the fact that one can interpret $\int\sqrt{g}R^2$ as a gauge fixing action which breaks the full Weyl group down to the harmonic Weyl subgroup, an operation known as partial gauge fixing \cite{Oda:2020wdd}.

In general, one could \emph{define} the harmonic Weyl group in arbitrary $d$
as the subset of Weyl transformations under which $R$ transforms homogeneously,
and this would be an alternative strategy to ``find'' it,\footnote{This leads to a generalized strategy discussed in Section~\ref{sect:generalization}.}
although in Sect.~\ref{sect:constructions} we adopt a different one based on the construction of the function $H_g(\Omega)$ from consistency requirements.
Whether the harmonic Weyl group is trivial or not on a given manifold depends on the existence
of solutions to the harmonic constraint $H_g(\Omega)=0$ and its boundary conditions,
but, in general, it always contain
at least rigid scale transformations (i.e.~homotheties), which trivially satisfy
$\Box_g \Omega=0$ in $d=4$ by virtue of $\Omega$ being constant, assuming the compatibility of the boundary conditions.

The harmonic Weyl transformations can be both finite or infinitesimal (in the sense of an $\Omega$ arbitrarily close to unity).
It might be illuminating to look at the structure of the infinitesimal restricted Weyl transformations, for simplicity on a $d=4$ dimensional manifold equipped with a metric $g_{\mu\nu}$ (i.e., when the constraint is linear). A first step of transformation can be carried out by an infinitesimal  Weyl transformation with a factor $\Omega_1=1+\epsilon \,\omega_1$ such that $\Box \, \omega_1=0$ with  $\omega_1$ finite and $\epsilon$ an arbitrarily small constant parameter. A subsequent second transformation, with factor $\Omega_2$, must satisfy
\begin{equation}
\label{second_step}
0= \Box_{\Omega_1^2 g}\Omega_2= \frac{1}{\Omega_1^2} \Bigl( \Box_g \Omega_2 +2 g^{\mu\nu} \partial_\mu (\ln{ \Omega_1 })\partial_\nu \Omega_2 \Bigr)
\end{equation}
and one can easily see that, perturbatively, a solutions arbitrarily close to the identity exist and can be written as
\begin{equation}
\Omega_2=1+\epsilon \, \Omega_2^{(1)}+ \epsilon^2 \,\Omega_2^{(2)} +\cdots \,,
\end{equation}
where $\Omega_2^{(1)}$ is an arbitrary finite harmonic function ($\Box_g \Omega_2^{(1)}=0$) and $\Omega_2^{(2)}$ is a solution of the second order perturbative term of Eq.~\eqref{second_step}, i.e.,~$\Box_g \Omega_2^{(2)}+ 2 g^{\mu\nu} \partial_\mu \omega_1 \partial_\nu \Omega_2^{(2)} =0$. This procedure can be iterated to any order in $\epsilon$.
This perturbative construction can be carried out also for general dimensions $d\neq 4$, although it becomes more involved due to the non linearity of Eq.~\eqref{gen_harmonic}. Indeed, to find a perturbative infinitesimal solution of Eq.~\eqref{gen_harmonic}, one needs to assume the existence of the series $\Omega_1=1+\Omega_1^{(1)}+\epsilon^2 \, \Omega_2^{(2)}+\cdots $ and solve order by order in $\epsilon$. However, non-linearities are unavoidable and affect the form of the $\Omega_2^{(i)}$ for $i \ge 2$.

That said, it is nevertheless useful to consider the linearization of
$\Omega(x)=1+\sigma(x)$, for some $\sigma \ll 1$, and correct the shortcoming later. In this case, we obtain from $ H_g(\Omega)$ that, at first order and in any dimension $d$, the linearized condition becomes $\Box_g \sigma=0$.
An amusing fact, which we exploit in Sect.~\ref{sect:harmonic-implications} is that it is possible to rewrite the complete (nonlinearized) condition $H_g(\Omega)=0$ in a form
which is analog to the linearized one, i.e., with a Laplacian acting on a powerlaw function of
$\Omega$ instead of $\Omega$ itself. In fact, with some straightforward manipulation, we can rewrite
\begin{equation}\label{eq:harmonic-pseudoinfinitesimal}
 H_g(\Omega)= \frac{2}{d-2} \Omega^{\frac{4-d}{2}} \Box_g \Omega^{\frac{d-2}{2}}\,.
\end{equation}
This rewriting highlights the fact that the cases $d=2$ and $d=4$ are special.
Obviously, in $d=4$ the harmonic condition itself becomes linear. The reason why $d=2$ is special is less clear from the above formula and might require a more in-depth analysis, but it is related to the fact that, in the limit $d=2$, the rewriting \eqref{eq:harmonic-pseudoinfinitesimal} becomes such that $H_g(\Omega) \sim \Box_g \log\Omega = \Box_g\sigma$ and one expects an infinite dimensional group of conformal isometries in flat spacetime. Later we concentrate in general on the case $d>2$ and, more specifically, $d=4$.

We conclude with the observation that solving Eq.~\eqref{eq:harmonic-pseudoinfinitesimal} for an infinitesimal rescaling provides a resummation to all orders in the same limit of the perturbative solution of Eq.~\eqref{gen_harmonic}.
Indeed, one may take advantage of the form of the constraint in Eq.~\eqref{eq:harmonic-pseudoinfinitesimal} to linearize $\Omega^{\frac{d-2}{2}}$ instead of $\Omega$. Solving it with a harmonic function $\omega$ (that is, $\Box_g \omega=0$), we have
\begin{equation}
\label{exact_expansion}
\Omega^{\frac{d-2}{2}} = 1+ \frac{d-2}{2}  \omega \quad  \Rightarrow \quad \Omega= \left(1+ \frac{d-2}{2}
\omega\right)^{\frac{2}{d-2}}
=1+\omega-\frac{d-4}{2}\omega^2+\frac{(d-3)(d-4)}{12} \omega^3 + \cdots\,.
\end{equation}
It is straightforward to check that a perturbative solution of Eq.~\eqref{gen_harmonic} of the form
\begin{align}
\Omega=1+\epsilon \, \Omega^{(1)}+ \epsilon^2 \,\Omega^{(2)} +\epsilon^3 \,\Omega^{(3)}+\cdots\,,
\end{align}
for $\epsilon \, \Omega^{(1)} = \omega$, exactly reproduces 
the expansion of Eq.~\eqref{exact_expansion}.\footnote{In particular this is easily achieved using the relation $\Box_g \omega^n=n(n-1) \omega^{n-2} g^{\mu\nu} \partial_\mu \omega \partial_\nu \omega$, valid for any harmonic function $\omega$.}
The conclusion is that perturbation theory is a useful tool to analyze the action of the groupoid for both finite and infinitesimal transformations and, in general, the form of the restricted submanifold.

\subsection{The Liouville-Weyl subgroup}\label{sect:liouville-group}

A much less known Weyl groupoid, which we refer to as the \emph{Liouville-Weyl group}, in part borrowing the name from Ref.~\cite{Kuhnel-Rademacher}, is the one generated by transformations of the form
\begin{equation}\label{eq:liouville}
 g_{\mu\nu} \to g'_{\mu\nu}= \Omega^2 g_{\mu\nu}\,, \qquad {\rm for} \qquad 
 L_{g,\mu\nu}(\Omega)=0
 \,,
\end{equation}
where we have introduced the differential constraint
\begin{equation}\label{eq:liouville-condition}
 L_{g,\mu\nu}(\Omega)=\Bigl(\nabla_\mu \nabla_\nu - \frac{1}{d} g_{\mu\nu} \Box_g\Bigr) \Omega^{-1}=0\,.
\end{equation}
For convenience, we denote the covariantly ``transverse'' differential operator $P_{g,\mu\nu}=\nabla_\mu \nabla_\nu - \frac{1}{d} g_{\mu\nu} \Box_g$, so that $L_{g,\mu\nu}(\Omega) = P_{g,\mu\nu}(\Omega^{-1})$.
Similarly to the example of the harmonic group,
it is nontrivial to show that the Liouville-Weyl subgroup is actually closed under composition and associative.

In this case we need to prove that,
given the consecutive transformations
$\Omega_1$ and $\Omega_2$ which satisfy $L_{g,\mu\nu}(\Omega_1)=0$
and $L_{\Omega_1^2g,\mu\nu}(\Omega_2)=0$, then
we have that the composition $\Omega=\Omega_1 \cdot \Omega_2$ satisfies
$L_{g,\mu\nu}(\Omega)=0$. Indeed, with a bit of work it is possible to show
\begin{equation}
L_{g,\mu\nu}(\Omega)
=
\frac{1}{\Omega_2} \, L_{g,\mu\nu}(\Omega_1)
+
\frac{1}{\Omega_1} \, L_{\Omega_1^2g,\mu\nu}(\Omega_2) \, .
\end{equation}
Likewise the previous example, we have associativity and identity elements, but, in general, the naive inverse of an element does not belong to the Liouville-Weyl subset of Weyl transformations, implying that it is a partial groupoid. In fact, if $L_{g,\mu\nu}(\Omega)=0$, then $L_{g,\mu\nu}(\Omega^{-1}) \neq 0$, unless $\Omega$ is constant.

Notice that, in general, the condition
$L_{g,\mu\nu}(\Omega)=0$ in \eqref{eq:liouville} is truly a tensorial
relation and its trace is trivial because of the presence of the dimension dependent coefficient in the definition that makes it transverse. In other words, Eq.~$\eqref{eq:liouville}$ 
is ``orthogonal'' to
the harmonic condition. This implies that the harmonic and Liouville-Weyl subgroups are independent, though they intersect at least in the trivial constant elements (see also the discussion of the next section).
The Liouville-Weyl condition, which has the form of a traceless symmetric two index tensor, is realized in general by $d(d+1)/2 -1$ equations, which impose nontrivial constraints on the $d(d+1)/2$ components of the metric and on the scalar function $\Omega$. It should therefore be expected that it can be realized only on special families of manifolds \cite{Kuhnel-Rademacher}.

The Liouville-Weyl subgroup of the Weyl group can be understood ``naturally'' as the set of
Weyl transformations that leave the traceless part of the Ricci tensor
invariant. Defining it as $\tilde{R}_{\mu\nu} \equiv R_{\mu\nu}-\frac{1}{d}R g_{\mu\nu}$,
we have that, for general Weyl transformations,
\begin{equation}
\tilde{R}_{\mu\nu} \to \tilde{R}'_{\mu\nu} = \tilde{R}_{\mu\nu} + (d-2)\Omega L_{g,\mu\nu}(\Omega)
\end{equation}
so $\tilde{R}_{\mu\nu}$ does not change
under the Liouville-Weyl transformations that satisfy $L_{g,\mu\nu}(\Omega)=0$ \cite{Kuhnel-Rademacher},
much like the harmonic group is the structure that leaves $R$ invariant, apart from a homogeneous rescaling. The two dimensional limit of the above transformation is special because in
$d=2$ we have that, locally and in some chart, $R_{\mu\nu} =\frac{R}{2} g_{\mu\nu}$ so $\tilde{R}_{\mu\nu}$ is trivial.

It has been shown in Ref.~\cite{Kuhnel-Rademacher} that, in the case of geodesically complete manifolds, the Liouville-Weyl condition, seen as the special transformation for which the change of the Ricci scalar under Weyl rescaling is proportional to the metric itself, implies that {\it ``unless the conformal transformation is homothetic, the only possibilities are standard Riemannian spaces of constant sectional curvature and a particular warped product with a Ricci flat Riemannian manifold''}~\cite{Kuhnel-Rademacher}. Therefore, apart from a (constant) scale transformation, i.e.,~a homothety, there exist some nontrivial families of manifold and Weyl transformations associated to the restricted Weyl case.

Much like the example of the harmonic Weyl group, the Liouville-Weyl group is not truly a group since any transformation depends on the metric itself. 
Infinitesimal transformations can be costructed as described previously for the harmonic Weyl case.
Then, at linear order, we can take $\Omega(x)=1+\sigma(x)$ and $\sigma(x)\ll 1$, in which case
the linearized condition becomes $P_{g,\mu\nu} \sigma=0$, higher order corrections being computable in perturbation theory.

\subsection{Nonexistence of other rank-$2$ tensorial Weyl subgroups and rigid scale invariance}

Relatively recently, an attempt has been made in exploring the intersection
of harmonic and Liouville-Weyl transformations, under the name of \emph{special Weyl group} in the first draft of Ref.~\cite[\tt arXiv:v1]{Shaposhnikov:2022dou},
with the purpose of generalizing the conformal group to
an arbitrary (pseudo)Riemannian manifold in an anomaly free way (i.e., not as the group of conformal isometries of a given manifold, which is subject to the trace-anomaly). The idea
was an interesting endeavour because, in flat space, the combination of the two conditions, $H_g(\Omega)=0$ and $L_{g,\mu\nu}(\Omega)=0$, results in solutions of $\Omega$
that can be assimilated to the action of the generators of the conformal
group, including the special conformal transformations.
Unfortunately, on a general manifold with nonzero curvature tensor, it is straightforward to show that, together, the two conditions also imply $R_{\mu\nu}{}^\rho{}_\theta \partial_\rho\Omega=0$. Consequently a putative special transformation $\Omega$ can be at most a rigid scale transformation, unless the manifold is flat
as correctly amended in the published second version
of Ref.~\cite[\tt arXiv:v2]{Shaposhnikov:2022dou}. This is true even for a geodesically incomplete manifold, so the assumptions of Ref.~\cite{Kuhnel-Rademacher} do not circumvent this limitation.

From our point of view the original attempt of Ref.~\cite{Shaposhnikov:2022dou}, even if unsuccessful, is worth mentioning because it highlights that the conformal anomaly
cannot be bypassed by combining the two subgroups, unless we restrict our attention to the case of flat space, which is classically nonanomalous to begin with. It also justifies the considerations on the anomaly that we explore later
and, for us, was source of inspiration for the analysis of Sect.~\ref{sect:constructions}, which resulted in the classification of the groupoids presented there.

The analysis of the intersection of the harmonic and Liouville-Weyl subgroups
hints at the possibility that group-like substructures may be related to further independent tensorial conditions like the Liouville-Weyl one.
One could define the \emph{tensor-harmonic Weyl group} as the transformations
\begin{equation}\label{eq:tensor-harmonic}
\begin{split}
 &g_{\mu\nu} \to g'_{\mu\nu}= \Omega^2 g_{\mu\nu}\,, 
 \\
 &H_{g,\mu\nu}(\Omega) = \nabla_\mu \partial_\nu \Omega
 -\frac{2}{\Omega} \partial_\mu \Omega \partial_\nu \Omega
 +\frac{1}{2\Omega}g_{\mu\nu} g^{\alpha\beta}\partial_\alpha \Omega \partial_\beta \Omega
 =0 \,.
 \end{split}
\end{equation}
In perfect analogy with the other examples, one can show that the associativity condition is fulfilled. In fact,
setting $\Omega=\Omega_1\cdot \Omega_2$, one has
\begin{equation}
H_{g,\mu\nu}(\Omega)
=
{\Omega_2} \, H_{g,\mu\nu}(\Omega_1)
+
{\Omega_1} \, H_{\Omega_1^2g,\mu\nu}(\Omega_2) \, ,
\end{equation}
so that one is tempted to define a restricted groupoid based on this condition.
However, it is important to realize that the tensor-harmonic constraint can be splitted in trace and trace-free parts as 
\begin{equation}\label{eq:tensHarmSplitting}
H_{g,\mu\nu}(\Omega)
=
\frac{g_{\mu\nu}}{d} H_{g}(\Omega)
-
\Omega^2 L_{g,\mu\nu}(\Omega)
\, ,
\end{equation}
because ${H_{g,}}^{\mu}{}_{\mu}(\Omega) = H_{g}(\Omega)$ (the trace becomes the scalar harmonic constraint). Therefore, imposing $H_{g,\mu\nu}(\Omega)=0$, is equivalent to simultaneously enforce $H_{g}(\Omega)=0$ and $L_{g,\mu\nu}(\Omega)=0$, which yields only trivial solutions
as explained above.
This can be checked explicitly by inspecting the combination
\begin{align}
\nabla_\alpha H_{g, \mu\nu}-\nabla_\mu H_{g, \alpha\nu}=0\,,
\end{align}
and using the form of the constraint Eq.~\eqref{eq:tensor-harmonic} to lower the rank of the derivatives acting on $\Omega$, yielding $R_{\alpha\mu}{}^\rho{}_{\nu}\partial_\rho\Omega=0$, as anticipated.
However, the splitting of Eq.~\eqref{eq:tensHarmSplitting} is important for the next section discussing the uniqueness, in which we conclude that the only differential constraints with two derivatives that yields nontrivial results (flat manifold/scale transformations) are the Liouville-Weyl and harmonic-Weyl constraints.

Geometrically, the tensor-harmonic condition is, modulo a factor,
the transformation of the Schouten tensor, defined as
\begin{equation}\label{eq:schouten-def}
 K_{\mu\nu}=\frac{1}{(d-2)}\Bigl(R_{\mu\nu}-\frac{1}{2(d-1)}g_{\mu\nu} R\Bigr)\,.
\end{equation}
In fact, for general Weyl transformations, we have
\begin{equation}\label{eq:schouten-transformations}
 K_{\mu\nu} \to K'_{\mu\nu} = K_{\mu\nu}-\Omega^{-1} H_{g,\mu\nu}(\Omega)\,,
\end{equation}
implying that $K_{\mu\nu}$ is invariant under tensor-harmonic transformations. 
The Schouten tensor is sometimes known as the ``trace-adjusted'' Ricci tensor
and it is ubiquitous in conformal geometry \cite{Fefferman:2007rka}, so it should not be surprising that it appears here.
Its trace is the Ricci scalar modulo a dimension dependent factor,
which we denote
\begin{equation}\label{eq:J-def}
 {\cal J} \equiv K^{\mu}{}_\mu = \frac{1}{2(d-1)} R\,,
\end{equation}
and transforms homogeneously under both tensor-harmonic and harmonic transformations
\begin{equation}
\begin{split}
 {\cal J} \to {\cal J}' &= \Omega^{-2}\Bigl(
 {\cal J}
 -\Omega^{-1} g^{\mu\nu} H_{g,\mu\nu}(\Omega)
 \Bigr)=\Omega^{-2}\Bigl({\cal J}
 -\Omega^{-1}  H_{g}(\Omega)
 \Bigr)\,.
 \end{split}
\end{equation}
This is expected given that ${\cal J}$ is a scalar proportional to $R$ and the trace of the tensor-harmonic condition is just the harmonic condition.
Because of Eq.~\eqref{eq:schouten-transformations}, the tensor-harmonic groupoid could be defined as the set of transformations that leave
$K_{\mu\nu}$ invariant. 
Indeed the decomposition given in Eq.~\eqref{eq:tensHarmSplitting} reflects precisely the rewriting of the Schouten tensor as a linear combination of the Ricci traceless tensor and the Ricci scalar.

\section{A classification of the groupoids of the Weyl group}\label{sect:constructions}

The natural question at this stage is to ask whether the three substructures (one of which is trivial)
discussed in the previous section are unique or not.
To avoid unnecessary formalism, we clarify the notions of generalization and uniqueness along the way. 
This analysis is based on the study of the constraints with two derivatives which, in general, are not necessarily derived from a Weyl variation of some geometric tensor. We find that there exists only another subgroup of the Weyl group constrained by a lightcone condition.
We will come back to the problem of generating Weyl substructures imposing conditions with an arbitrary number of derivatives in Section~\ref{sect:generalization}, where we instead restrict to the constraints obtained by Weyl transformations of suitable tensors,
since this approach is computationally more convenient.

\subsection{Scalar differential constraints}

To frame the discussion in more concrete terms,
we concentrate first on potential generalizations of the harmonic condition given in \eqref{eq:harmonic}, i.e., substructures defined by a differential constraint based on a scalar functional differential equation that depends only on covariant derivatives of the conformal factor $\Omega$.
We can parametrize the most general scalar condition constructed with up to
two derivatives acting on $\Omega$ as a homogeneous (of degree two) equation introducing two constant scalar parameters $s_i$ as
\begin{equation}\label{eq:general-scalar-2d}
 C_g(\Omega) = s_1 \Omega \Box_g \Omega
 + s_2 g^{\alpha\beta}\partial_\alpha \Omega \partial_\beta \Omega\,,
\end{equation}
which reproduces $H_g(\Omega)$ for a special choice of $s_i$ if divided
by the conformal factor itself. We also require that $\Omega \neq 0$, so that the $\Omega$s are nondegenerate and part of the Weyl group.
Notice that, from the ansatz \eqref{eq:general-scalar-2d}, we have momentarily excluded a possible term that does depend only on $\Omega^2$, e.g., $R \Omega^2$,
so the equation \eqref{eq:general-scalar-2d} is also shift-invariant, for constant shifts of $\Omega \to \Omega + c$. Shift-invariance ensures that $\Omega=1$ is always a solution, but we shall discuss the more general case later in this section.

Let us look for consistent groupoids of the Weyl group defined by $C_g(\Omega)=0$, for $C_g(\Omega)$ given in \eqref{eq:general-scalar-2d}. The consistency of the groupoid should determine the values of the parameters $s_i$.
We consider first the case $s_1\neq 0$, so we can safely take $s_1=1$ without loss of generality being the equation homogeneous.
In order for $C_g(\Omega)=0$ to be the condition of a
subgroup of the Weyl group, the combination of two consecutive transformations $\Omega_1$ and $\Omega_2$, constrained appropriately,
must obey the same constraint as the original one.
In practice, if $C_g(\Omega_1)=0$ and $C_{\Omega_1^2 g}(\Omega_2)=0$,
we must also have unequivocally that $C_g(\Omega_1\cdot \Omega_2)=0$.

First consider $C_g(\Omega_1)=0$, for which, using \eqref{eq:general-scalar-2d} with the replacement $\Omega \to \Omega_1$, we find the solution in terms of $\Box_g \Omega_1$ (recall $s_1=1$) as
\begin{equation}
 \Box_g \Omega_1=-s_2\Omega_1^{-1}\partial^\mu\Omega_1\partial_\mu\Omega_1\,,
\end{equation}
as a result of the constraint on the first transformation. Then consider $C_{\Omega_1^2 g}(\Omega_2)=0$,
for which we need the expression in terms of the previous metric $g_{\mu\nu}$
\begin{equation}
 C_{\Omega_1^2 g}(\Omega_2)=\Omega_1^{-2}\Omega_2\Bigl\{ \Box_g \Omega_2+s_2\Omega_2^{-1}\partial^\mu\Omega_2\partial_\mu\Omega_2+(d-2)\Omega_1^{-1}\partial^\mu \Omega_1 \partial_\mu\Omega_2
 \Bigr\}
 \,.
\end{equation}
We find the solution of the second constraint in terms of $\Box_g \Omega_2$ as
\begin{equation}
 \Box_g \Omega_2=-s_2\Omega_2^{-1}\partial^\mu\Omega_2\partial_\mu\Omega_2-(d-2)\Omega_1^{-1}\partial^\mu \Omega_1 \partial_\mu\Omega_2
 \,.
\end{equation}
We can then insert the solutions of $\Box_g \Omega_1$ and $\Box_g \Omega_2$ in the expression of $C_g(\Omega_1\cdot \Omega_2)$, which again comes from \eqref{eq:general-scalar-2d} replacing $\Omega \to \Omega_1\cdot\Omega_2$ and distributing the covariant derivatives. This results in the elimination of the terms with $\Box_g \Omega_i$. The only surviving term is the one mixing the derivatives of $\Omega_1$ and $\Omega_2$
\begin{equation}
 C_g(\Omega_1 \cdot \Omega_2) = \bigl(2 s_2 - (d-4)\bigr)\partial^\mu \Omega_1 \partial_\mu\Omega_2 \qquad {\rm if} \quad C_g(\Omega_1)=C_{\Omega_1^2 g}(\Omega_2)=0 \,.
\end{equation}
The combined transformation is in general a solution only if $C_g(\Omega_1\cdot \Omega_2)=0$, so only if $s_2=\frac{d-4}{2}$ for general $\Omega_1$ and $\Omega_2$.
In fact, we can show that, in general and for $s_1=1$,
\begin{equation}
C_g(\Omega_1\cdot \Omega_2 )=\Omega^2_2 \, C_g(\Omega_1)+ \Omega^4_1 \, C_{\Omega_1^2 g}(\Omega_2) +\bigl(2 s_2 - (d-4)\bigr)\partial^\mu \Omega_1 \partial_\mu\Omega_2\,,
\end{equation}
from which it becomes trivial to establish that $C_g(\Omega_1\cdot \Omega_2 )=0$ if $C_g(\Omega_1)=C_{\Omega_1^2 g}(\Omega_2)=0$ and $s_2=\frac{d-4}{2}$, provided $\partial^\mu \Omega_1 \partial_\mu\Omega_2 \ne 0$ (see next subsection).
As a consequence,
we find that the scalar condition for which the associative property holds is
\begin{equation}\label{eq:scalar-solution1}
 C_g(\Omega) \to \Omega \Box_g \Omega
 + \frac{d-4}{2} g^{\alpha\beta}\partial_\alpha \Omega \partial_\beta \Omega = \Omega H_g(\Omega)\,,
\end{equation}
proportional to $H_g(\Omega)$, already defined in \eqref{eq:harmonic}.

\subsubsection{The special case $s_1=0$ and the lightcone}\label{lightcone_case}

We now consider the case $s_1=0$, which corresponds to the constraint
\begin{equation}\label{eq:scalar-solution2}
\bar{C}_g(\Omega)=g^{\alpha\beta}\partial_\alpha \Omega \partial_\beta \Omega
=0\,.
\end{equation}
It has the form of a lightcone condition for the gradient vector orthogonal to the null hypersurface of codimension $2$ which one obtains imposing both $\Omega(x)= c$ and $\bar{C}_g(\Omega)=0$ for $c$ constant.
We also need to impose the associativity condition, for which we first notice that $\bar{C}_{\Omega_1^2 g}(\Omega_2)=\Omega_1^{-2} \bar{C}_{ g}(\Omega_2)$ and therefore, because of the lightcone nature, the equation for $\Omega_2$ is the same as the one for $\Omega_1$. Then we must check that $\bar{C}_g(\Omega_1 \Omega_2)=0$ is preserved and this implies the crossing condition
$g^{\alpha\beta}\partial_\alpha \Omega_1 \partial_\beta \Omega_2=0$ which is satisfied for 
\begin{equation}\label{associativity_special}
\partial_\beta \Omega_2= \rho(x) \partial_\beta \Omega_1\,,
\end{equation}
namely the two gradients should be proportional up to an arbitrary function $\rho(x)$.
Since Eq.~\eqref{eq:scalar-solution2} is equivalent to the one constructing a geodesic along the gradient of $\Omega$, this means that the geodesics along the lightcone have a tangent vector given by the gradient of the expanding $\Omega$ along subsequent restricted Weyl transformations. These properties make this lightcone substructure an actual subgroup of the Weyl group (not a groupoid).

The lightcone condition $\bar{C}_g(\Omega)=0$ clearly can admit solutions depending on $g_{\mu\nu}$, so let us consider a couple of simple examples.
In a conformally flat manifold with $g_{\mu\nu}=A(x) \eta_{\mu\nu}$ the solutions are parameterized by a momentum $p_\mu$ such that $p^2=\eta^{\mu\nu} p_\mu p_\nu=0$ and are given by $\Omega_p(x)= f(p_\mu x^\mu)$ in terms of an arbitrary function $f$, so that $\partial_\alpha \Omega \propto p_\alpha$. For multiple transformations, infinitesimal or finite, one must remember that the solutions must be associated to the same light cone vector $p_\mu$ of the two linearly independent spanning the space of light-cone vectors, to respect the associativity requirement.
Another simple case can be found for spherical symmetry, with a static metric line element $d s^2=-A(r) dt^2+B(r) dr^2+r^2 (d\theta^2+\sin^2\theta d\phi^2)$, in which case the constraint for a spherically symmetric $\Omega(t,r)$ reads
\begin{equation}\label{eq:scalar-solution3}
\bar{C}_g(\Omega)=-\frac{1}{A}\left[ \left( \partial_t \Omega \right)^2 -\frac{A(r)}{B(r)}  \left( \partial_r \Omega \right)^2  \right] = 0 \,  \quad
\Rightarrow \quad  \left( \partial_t   \mp \partial_{\tilde{r} }\right) \Omega (t, \tilde{r} ) = 0 \, ,
\end{equation}
where $\frac{d\tilde{r}}{d r}=\left(\frac{B}{A}\right)^{1/2}$ and the solutions are given by $\Omega_\pm(t,r) =f(t \pm \tilde{r}(r) )$ with $f$ an arbitrary function. 
Finally, we have to remember in the composition of trasformations to use the solutions with a gradient proportional, as required in Eq.~\eqref{associativity_special}, i.e., compose solutions $\Omega_i$ with the same sign $i=\pm$.
The equation should be eventually supplied with boundary conditions.

We note that the $\bar{C}_g(\Omega)$ condition can possibly be imposed in conjunction with the Harmonic condition, leading in any dimension to $\Box_g \Omega=0$ together with $\bar{C}_g(\Omega)=0$.\footnote{Note that $\Omega$ is not a dynamical scalar. For a free scalar field with a two derivative kinetic term the equation of motion $\Box_g\phi=0$ differs from $\partial_\mu \phi \partial^\mu \phi=0$ by a boundary term.}
We finally anticipate that the quadratic lightcone condition $\bar{C}_g(\Omega)=0$ cannot be obtained from requiring a Weyl transformation invariance/covariance of some curvature scalar that depends on two derivatives acting on the metric. The only potential candidate would be the Ricci scalar, which leads to the harmonic condition.

\subsubsection{Scalar constraint with a homogeneous nonderivative term}

One might wonder why we have not included a nonderivative term to \eqref{eq:general-scalar-2d}. As an exercise, we generalize the constraint \eqref{eq:general-scalar-2d} by including an additional term $s_3 R \Omega^2$.
Replacing \eqref{eq:general-scalar-2d} with
\begin{equation}\label{eq:general-scalar-2d-nonhomogeneous}
 C_g(\Omega) = s_1 \Omega \Box_g \Omega
 + s_2 g^{\alpha\beta}\partial_\alpha \Omega \partial_\beta \Omega
 +s_3 R \Omega^2\,.
\end{equation}
If $s_3 \neq 0$, the additional term removes the ``identity'' solution $\Omega=1$.
We can repeat the same steps as the previous case for $s_1=1$. We still have the solution \eqref{eq:scalar-solution1} for $s_3=0$, but we also have a second independent solution
\begin{equation}\label{eq:scalar-solution2-2}
 C_g(\Omega) \to \Omega \Box_g \Omega + \frac{d-4}{2}\partial_\mu \Omega \partial^\mu \Omega
 -\frac{1}{2(d-1)} R \Omega^2
 \,.
\end{equation}
The meaning of the above equation can be clarified if we rewrite it as
\begin{equation}
 C_g(\Omega) = -\frac{1}{2(d-1)} \Omega^4 R |_{\Omega^2 g}
 = -\Omega^4 {\cal J}|_{\Omega^2 g}
 \,,
\end{equation}
where $R |_{\Omega^2 g}$ is the scalar curvature of the metric $g'_{\mu\nu}=\Omega^2 g_{\mu\nu}$. Recall that ${\cal J}$ is just $R$ up to a dimension dependent constant, as defined in \eqref{eq:J-def}. Therefore the inclusion of the nonderivative term to 
\eqref{eq:general-scalar-2d} has the only effect of displaying a groupoid which is the semigroupoid of conformal transformations that map a metric $g_{\mu\nu}$ to one for which the scalar curvature is zero,
i.e., they map to a Ricci-scalar-flat metric. The case $s_1=0$ does not reveal additional solutions. We have not found interesting applications to these types of groupoids, although we do not exclude that they exist, so in the rest of the paper we only concentrate on differential constraints that are shift-symmetric in $\Omega$, which are guaranteed to have $\Omega=1$ as solution and thus a set of identity elements (see also appendix \ref{sect:appendix-groupoids}).

\subsection{Tensor constraints}

The procedure used to construct the substructures of the Weyl group starting from  the general scalar constraint \eqref{eq:general-scalar-2d}
can be generalized considering 
the most general two-index tensor condition in the general parametric form
\begin{equation}\label{eq:2d-tensor-ansatz}
C_{g,\mu\nu}(\Omega)=c_1 \Omega\nabla_\mu\partial_\nu \Omega
 +c_2 g_{\mu\nu} \Omega \Box_g \Omega
 +c_3 \partial_\mu\Omega\partial_\nu\Omega
 +c_4 g_{\mu\nu} g^{\alpha\beta}\partial_\alpha \Omega \partial_\beta \Omega\,,
\end{equation}
for some choices of the parameters $c_i$. 
The above ansatz includes all possible (homogeneous of degree two) terms with two uncontracted indices
that are obtained by acting with two derivatives on the conformal factors $\Omega$. 
It is also symmetric in the two indices because of the symmetry of the connection, $\nabla_\mu\partial_\nu \Omega = \nabla_\nu\partial_\mu \Omega$.

As a first step we choose to divide the analysis based on the value of $c_1$, corresponding to $c_1=0$ and $c_1\ne 0$.
Let us start from the case $c_1=0$. One can consider the trace and traceless parts of~\eqref{eq:2d-tensor-ansatz} which read
\begin{eqnarray}
&{}& c_2 \Omega \Box_g{\Omega}+ \left(c_4+\frac{c_3}{d}\right) g^{\alpha\beta}\partial_\alpha \Omega \partial_\beta \Omega =0\,, \nonumber\\
&{}& c_3\left( \partial_\mu\Omega\partial_\nu\Omega -\frac{1}{d} g_{\mu\nu} g^{\alpha\beta}\partial_\alpha \Omega \partial_\beta \Omega \right) =0 \,,
\end{eqnarray}
respectively.
Taking into account the fact that in the second equation the first term is a matrix with zero determinant while the matrix in the second term is not, one can easily see that no general solutions exist besides the $\Omega$ constant case. Therefore, for a nontrivial solution to exist, one must have $c_3=0$,
which gives a scalar constraint $C_{g,\mu\nu}(\Omega) = g_{\mu\nu} C_g(\Omega)$ if we identify $s_1=c_2$ and $s_2=c_4$ when comparing \eqref{eq:general-scalar-2d} with \eqref{eq:2d-tensor-ansatz}. This case thus reduces to the general scalar constraint discussed in the previous section because the metric is by definition nondegenerate. Therefore this case has already been covered in the previous section, and we can ignore it here.

Moving on to the case $c_1\neq 0$,
we can set $c_1=1$ by rescaling the complete expression without loss of generality. 
Notice that $C_{g,\mu\nu}(\Omega)=0$
can be solved tensorially as 
\begin{equation}\label{eq:tensor-step-1}
 \Omega\nabla_\mu\partial_\nu \Omega
 =
 -c_2 g_{\mu\nu} \Omega \Box_g \Omega
 -c_3 \partial_\mu\Omega\partial_\nu\Omega
 -c_4 g_{\mu\nu} g^{\alpha\beta}\partial_\alpha \Omega \partial_\beta \Omega\,.
\end{equation}
The simplest trick to find the consistent choices of the $c_i$
is to take the trace of Eq.~\eqref{eq:tensor-step-1}, which gives an expression for $\Omega \Box_g\Omega$, and substitute it back into the monomial $\Omega \Box_g \Omega$ that appears on the right hand side of Eq.~\eqref{eq:tensor-step-1} itself.
With this procedure we find
\begin{equation}\label{eq:tensor-step-2}
 \Omega\nabla_\mu\partial_\nu \Omega
 = d c_2^2 g_{\mu\nu} \Omega \Box_g \Omega
 -c_3 \partial_\mu\Omega\partial_\nu\Omega
 +(c_2 c_3- c_4+c_2 c_4 d)g_{\mu\nu} g^{\alpha\beta}\partial_\alpha \Omega \partial_\beta \Omega\,.
\end{equation}
If we require that the two equations are consistent, i.e., that the right hand sides of Eqs.~\eqref{eq:tensor-step-1} and \eqref{eq:tensor-step-2} coincide, we have two general solutions,
either $c_2=0$, which eliminates $\Box_g \Omega$ altogether from the right hand side of \eqref{eq:tensor-step-1},
or $c_2=-\frac{1}{d}$ and $c_4=-\frac{c_3}{d}$.
The important distinction is in the different values of $c_2$ and thus we have two separate cases to consider.

\subsubsection{Case $c_2=0$: the tensor-harmonic subgroup}

As first case of interest, we consider $c_2=0$ in \eqref{eq:2d-tensor-ansatz}.  We start by taking $c_2=0$ in \eqref{eq:2d-tensor-ansatz}, which becomes
\begin{equation}
 C_{g,\mu\nu}(\Omega) \to 
 \Omega\nabla_\mu\partial_\nu \Omega
 +c_3 \partial_\mu\Omega\partial_\nu\Omega
 +c_4 g_{\mu\nu} g^{\alpha\beta}\partial_\alpha \Omega \partial_\beta \Omega
 \,,
\end{equation}
and both $c_3$ and $c_4$ are still undetermined.
In order to form a groupoid of the Weyl group we must require that two consecutive transformations $\Omega_1$ and $\Omega_2$ satisfy the original relation, and the procedure is the same as the previous section: $C_{g,\mu\nu}(\Omega_1)=0$ and $C_{\Omega_1^2 g,\mu\nu}(\Omega_2)=0$ must imply $C_{g,\mu\nu}(\Omega_1\Omega_2)=0$.

We solve $C_{g,\mu\nu}(\Omega_1)=0$ in terms of $\nabla_\mu\partial_\nu \Omega_1$, to obtain 
\begin{equation}
\nabla_\mu\partial_\nu \Omega_1
=
-\Omega^{-1}_1
\Bigl(
c_3 \partial_\mu \Omega_1 \partial_\nu \Omega_1
+
c_4 g_{\mu\nu} g^{\alpha\beta}\partial_\alpha \Omega_1 \partial_\beta \Omega_1
\Bigr)
\, ,
\end{equation}
Then we solve $C_{\Omega_1^2 g,\mu\nu}(\Omega_2)=0$ in terms of $\nabla_\mu\partial_\nu \Omega_2$ and using the metric $g_{\mu\nu}$, to obtain
\begin{equation}
\nabla_\mu\partial_\nu \Omega_1
=
\Omega^{-1}_1
\Bigl(
2 \partial_{(\mu} \Omega_1 \partial_{\nu)} \Omega_2
-
 g_{\mu\nu} g^{\alpha\beta}\partial_\alpha \Omega_1 \partial_\beta \Omega_2
\Bigr)
-
\Omega^{-1}_2
\Bigl(
c_3 \partial_\mu \Omega_2 \partial_\nu \Omega_2
+
c_4 g_{\mu\nu} g^{\alpha\beta}\partial_\alpha \Omega_2 \partial_\beta \Omega_2
\Bigr)
\, .
\end{equation}
As in the previous section, we expressed both solutions in terms of the covariant derivative $\nabla_\mu$, which is compatible only with $g_{\mu\nu}$.
Substituting the previous two equations in $C_{g,\mu\nu}(\Omega_1\cdot \Omega_2)$, a rather long, but straightforward, computation shows
\begin{equation}
C_{g,\mu\nu}(\Omega_1\cdot\Omega_2)
=
 \Omega_1 \Omega_2
\Bigl(
2(2+c_3)
 \, \partial_{(\mu} \Omega_1 \partial_{\nu)} \Omega_2
+
(2c_4-1) g_{\mu\nu} \, \partial_\alpha \Omega_1 \partial_\beta \Omega_2
\Bigr)
\, ,
\end{equation}
that depends only on monomials mixing the $\Omega_i$s.
Therefore $C_{g,\mu\nu}(\Omega_1\cdot\Omega_2)=0$ holds only for $c_3=-2$ and $c_4=\frac{1}{2}$, which eliminate the tensors mixing derivatives of the $\Omega_i$s. This implies that
the closure property holds for transformations constrained by 
\begin{equation}
 C_{g,\mu\nu}(\Omega) \to 
 \Omega\nabla_\mu\partial_\nu \Omega
 -2 \partial_\mu\Omega\partial_\nu\Omega
 +\frac{1}{2} g_{\mu\nu} g^{\alpha\beta}\partial_\alpha \Omega \partial_\beta \Omega\,.
\end{equation}
This differential constraint can be recasted in the form of the tensorial condition
\eqref{eq:tensor-harmonic} of the tensor-harmonic Weyl group,
i.e. $C_{g,\mu\nu}(\Omega)= \Omega H_{g,\mu\nu}(\Omega)$, which nevertheless, as previously discussed, leads to a trivial condition and is of no interest.

\subsubsection{Case $c_2=-\frac{1}{d}$: the Liouville-Weyl subgroup}

The second case corresponds to
$c_2=-\frac{1}{d}$ and $c_4=-\frac{c_3}{d}$ in \eqref{eq:2d-tensor-ansatz}. We have thus that
\begin{equation}
C_{g,\mu\nu}(\Omega) \to
 \Omega\nabla_\mu\partial_\nu \Omega
 -\frac{1}{d} g_{\mu\nu} \Omega \Box_g \Omega
 +c_3 \Bigl(\partial_\mu\Omega\partial_\nu\Omega
 -\frac{1}{d} g_{\mu\nu} g^{\alpha\beta}\partial_\alpha \Omega \partial_\beta \Omega
 \Bigr)\,.
\end{equation}
To impose associativity we follow exactly the same strategy as the previous case, i.e., solving in $\nabla_\mu \partial_\nu \Omega_i$ the constraints on the $\Omega_i$.
In this case, we find that we must require $c_3=-2$.
The groupoid property thus implies
\begin{equation}
 C_{g,\mu\nu}(\Omega) \to
 \Omega\nabla_\mu\partial_\nu \Omega
 -\frac{1}{d} g_{\mu\nu} \Omega \Box_g \Omega
 -2 \Bigl(\partial_\mu\Omega\partial_\nu\Omega
 -\frac{1}{d} g_{\mu\nu} g^{\alpha\beta}\partial_\alpha \Omega \partial_\beta \Omega
 \Bigr)\,,
\end{equation}
that can be cast in the form of the constraint \eqref{eq:liouville} of the Liouville-Weyl subgroup. The simplest way to see this is to expand \eqref{eq:liouville} so derivatives act on $\Omega$, rather than $\Omega^{-1}$, and multiply the final result by the overall factor $\Omega^3$,
so that
$C_{g,\mu\nu}(\Omega)=- \Omega^3 L_{g,\mu\nu}(\Omega)$.

\subsubsection{Final remarks on the construction and uniqueness}

We can summarize our findings for the non trivial tensor conditions ($c_1\ne 0$) corresponding to the cases $c_2=0$ and $c_2=-\frac{1}{d}$ as 
\begin{equation}
 C_{g,\mu\nu}(\Omega) =\Omega H_{g,\mu\nu}(\Omega)+c_2 \,g_{\mu\nu} \Omega H_g(\Omega) \,,
 \end{equation}
 where $H_{g,\mu\nu}(\Omega)$ satisfies the relation~\eqref{eq:tensHarmSplitting}.

Another way to distinguish among these two cases
is to realize that the requirement of consistency between \eqref{eq:tensor-step-1} 
and \eqref{eq:tensor-step-2} simply selects the traceful vs traceless versions of the solutions of $\nabla_\mu \partial_\nu \Omega$ and of the ansatz $C_{g,\mu\nu}(\Omega)$ in \eqref{eq:2d-tensor-ansatz} itself. So another way to go about to find the two cases is to separate \eqref{eq:2d-tensor-ansatz} in the traceful case, for which the trace $\Box_g\Omega$ is determined and obviously must give the harmonic condition to be substituted back into the constraint itself
and in the traceless case, for which $\Box_g \Omega$ is undetermined by the original equation.
Following this alternative strategy, we once again return to the two cases $c_2=0$ or $c_2=-\frac{1}{d}$ discussed in the previous two sections, respectively.

In the ansatz \eqref{eq:2d-tensor-ansatz} we have also neglected terms of the form $g_{\mu\nu} R \Omega^2$ and $R_{\mu\nu}\Omega^2$ (non-shift invariant) for the same reasons explained in the scalar case. There are, in fact, additional group-like substructures (specifically semigroupoids, given that $\Omega =1$ is not a solution) if these two terms are included.

In conclusion, combining all the results of the previous sections,
we are lead to the implication that the harmonic, lightcone and Liouville-Weyl 
subgroups are the \emph{unique} non trivial possibilities when restricting our attention to
subgroups of the Weyl group which satisfy (shift-invariant) conditions with up to two derivatives
acting on the conformal factor $\Omega$.
For generalizations with four and six derivatives of the scalar case see Sect.~\ref{sect:generalization}.

\section{Groupoids of Weyl group and the conformal anomaly}
\label{sect:implications-and-anomaly}

Now we are concerned with physical implications of the invariance of matter  field actions under the substructures of the Weyl group. The effect of full Weyl invariance has been discussed briefly at the beginning of Sect.~\ref{sect:intro}
and now we generalize that discussion to two groupoids substructures.
Here we discuss the implications concentrating on some examples, while a general strategy for constructing actions invariant under the non trivial groupoids summarized in Sect.~\ref{sect:intro}
is given in Sect.~\ref{sect:general-recipe}.

\subsection{Implications of harmonic Weyl invariance}
\label{sect:harmonic-implications}

Consider a covariant classical action $S[\Psi,g]$ of a general field $\Psi$ coupled to the metric. We want to explore the implications of the invariance of $S[\Psi,g]$ under the harmonic Weyl subgroup introduced in Sect.~\ref{sect:harmonic-group}. The action is, in general, not invariant under the full Weyl group. We choose the field $\Psi$ to transform as $\Psi \to \Omega^{w_\Psi} \Psi$ for some weight $w_\Psi$
under the action of the Weyl group and its subgroups. The energy-momentum tensor is defined $T^{\mu\nu}=-\frac{2}{\sqrt{g}} \frac{\delta S}{\delta g_{\mu\nu}}$
as in \eqref{eq:emt-weyl}, and it is conserved on-shell, $\nabla_\mu T^{\mu\nu}=0$, as a consequence of diffeomorphism invariance of $S[\Psi,g]$,
for which the transformations are $\delta_\xi g_{\mu\nu} = {\cal L}_\xi g_{\mu\nu}$ and $\delta_\xi \Psi = {\cal L}_\xi \Psi$ (${\cal L}$ is the Lie-derivative and $\xi^\mu$ is the vector field generating the infinitesimal diffeomorphism).
We assume that diffeomorphism invariance is nonanomalous, meaning that any anomaly, if present, is in the Weyl transformations.

Suppose, for a moment, that the action $S[\Phi,g]$ was invariant under the full Weyl group. This would imply that, solving on-shell $\frac{\delta S}{\delta \Psi}=0$, the infinitesimal Weyl transformation
\begin{equation}
 \delta_\sigma S[\Psi,g] = - \int {\rm d}^dx \sqrt{g} T^\mu{}_\mu \sigma\,,
\end{equation}
would be zero, for $\Omega=1+\sigma$ and general $\sigma\ll 1$. This would imply
that the energy-momentum tensor would be traceless.
However, given that we choose $S[\Psi,g]$ to be invariant only under the harmonic subgroup, we must have that, for a generic $\sigma$, the right hand side differs from zero precisely by a term that, instead, goes to zero with the harmonic condition, i.e.,
\begin{equation}\label{eq:harmonic-implication-mistake}
 \int {\rm d}^dx \sqrt{g} \bigl(
 T^\mu{}_\mu \sigma
 - \Phi \, \Box_g \sigma
 \bigr)
 =0\,,
\end{equation}
where we have introduced a scalar operator $\Phi$ that depends, in general, on $\Psi$ and its derivatives, so $\Phi=\Phi[\Psi]$ (or some renormalized counterpart in the quantized case).

We stress that, even if we do not retain the nonlinear terms of the Harmonic condition, this is sufficient because the energy-momentum tensor appears in a first order variation.
In fact, it is possible to replace in \eqref{eq:harmonic-implication-mistake} the infinitesimal difference with the actual finite difference and the infinitesimal constraint with the finite one in $d>2$  (the $d=2$ case should be treated separately) 
\begin{equation}\label{eq:harmonic-implication-correct}
\begin{split}
 \delta_\sigma S[\Psi,g] \to 
 S[\Omega^{w_\Psi}\Psi,\Omega^2 g]-S[\Psi,g]\,,
 \qquad \qquad
 \Box_g \sigma \to \frac{2}{d-2} \Omega^{\frac{4-d}{2}} \Box_g \Omega^{\frac{d-2}{2}}\,,
\end{split}
\end{equation}

Going back to \eqref{eq:harmonic-implication-mistake} for simplicity,
the requirement $\delta_\sigma S[\Psi,g]=0$ thus implies in general
that the trace of the EMT of a harmonic Weyl invariant theory is the Laplacian of some scalar operator
\begin{equation}\label{eq:harmonic-implication-T}
 T^\mu{}_\mu= \Box_g \Phi\,.
\end{equation}
The scalar operator has dimension $d-2$ and must be constructed from the field $\Psi$ and its covariant derivatives, so in general $\Phi =\Phi[\Psi]$,
but it has no ``operative'' definition (e.g., it is not defined as a functional derivative of the action with respect to some source), but rather it is defined indirectly from Eq.~\eqref{eq:harmonic-implication-T} itself.\footnote{
Operatively, $T_{\mu\nu}$ is the functional derivative of the action with respect to the metric.
An example of operative definition for the right hand side of \eqref{eq:harmonic-implication-T} would be the one used for the virial current in Ref.~\cite{Zanusso:2023vkn}, whose divergence replaces $\Box_g \Phi$ in \eqref{eq:harmonic-implication-T}. The virial current is then operatively defined as the functional derivative of the action with respect to the dilatation gauge potential.
}
This condition on the trace of the EMT is stronger than simple
scale invariance, which would imply that $T^\mu{}_\mu$ is the divergence of a current known as the virial current.
In fact, a $T_{\mu\nu}$ satisfying \eqref{eq:harmonic-implication-T}
can always be ``improved'' to a traceless form. In flat space, where the metric is $g_{\mu\nu}=\eta_{\mu\nu}$, such improvement would be
\begin{equation}\label{eq:T-improvement-R}
 T^{\mu\nu} \to T'^{\mu\nu}=T^{\mu\nu} + \frac{1}{d-1}(\partial^\mu\partial^\nu-g^{\mu\nu}\Box_\eta)\Phi
\end{equation}
which gives a \emph{different} EMT such that $T'{}^\mu{}_\mu=0$.
In order to interpret $T'^{\mu\nu}$
as (the flat space limit of) a variational energy momentum tensor of a \emph{different} action, the original action
must be modified by including a new term,
$S[\Psi,g] \to S'[\Psi,g]=S[\Psi,g]-\Delta S[\Psi,g]$,
where
\begin{equation}
 \Delta S[\Psi,g] = \frac{1}{2(d-1)}\int {\rm d}^d x \sqrt{g} \Phi R
 = \int {\rm d}^d x \sqrt{g} \Phi {\cal J}\,,
\end{equation}
because the functional variation with respect to the metric of ${\cal J}$
conveniently produces the second term on the right hand side of \eqref{eq:T-improvement-R}. 
The action $S'[\Psi,g]$ enjoys full Weyl invariance by construction.

In practice, this means that a harmonic Weyl invariant action
always differs from a fully Weyl invariant one by the inclusion of a term that couples the scalar curvature $R$ (or the conveniently normalized counterpart ${\cal J}=\frac{1}{2(d-1)} R$) with the scalar that can be integrated from the EMT in \eqref{eq:harmonic-implication-T}. We stress that the process of ``improving'' the action to full Weyl invariance does change the action itself as a field theory in curved space, in fact $S'$ might have less freedom in terms of couplings as we see in the examples below, but the conformal field theories arising in the flat space limit are arguably the same.

\subsubsection{The simplest example: harmonic Weyl invariant simple scalar in $d=4$}\label{sect:restricted-example-1}

For the examples we concentrate first on the general $d$ case and then on the physically interesting case $d=4$ that admits self-interaction \cite{Edery:2014nha}.
The most general harmonic Weyl invariant action of a scalar $\varphi$ with weight $w_\varphi=\frac{2-d}{2}$ is
\begin{equation}
 S[\varphi,g] = \int {\rm d}^4 x \sqrt{g}\Bigl\{\frac{1}{2}\partial_\mu\varphi\partial^\mu\varphi + \frac{\xi}{2} R \varphi^2 \Bigr\}\,,
\end{equation}
and it is invariant for any value of the coupling $\xi$ and in any $d$, which can be seen easily because $\sqrt{g} g^{\mu\nu}$ and $\sqrt{g} R$ transform both homogeneously to compensate the scalar weight $w_\varphi$.
Concentrating on the case $d=4$, where $w_\varphi=-1$, we can add a self-interaction with coupling $\lambda$
\begin{equation}\label{dim4self}
 S[\varphi,g] = \int {\rm d}^4 x \sqrt{g}\Bigl\{\frac{1}{2}\partial_\mu\varphi\partial^\mu\varphi + \frac{\xi}{2} R \varphi^2 + \frac{1}{4!}\lambda \varphi^4\Bigr\}\,,
\end{equation}
which is invariant for any $\lambda$ up to boundary terms.
The above two actions can be derived directly, or, alternatively, using the methods described later in Sect.~\ref{sect:general-recipe}.
In the flat space limit the interaction with $R$ decouples and the action is conformally invariant.
As it is well-known,
the requirement of full classical Weyl invariance would fix the value of the coupling $\xi$ to be $\frac{d-2}{4(d-1)}$ in general $d$ and $\frac{1}{6}$ in $d=4$. Quantum effects would have implications also for the coupling $\lambda$ as a consequence of the conformal anomaly, but classically $\lambda$ can take any value. In general, for arbitrary $\xi$, the action is manifestly only scale invariant in curved spacetime by dimensional analysis.

Using the variational definition \eqref{eq:emt-weyl}, we now concentrate on the case $d=4$ and compute the energy momentum tensor off-shell (however, everything discussed in this section works for general $d$ as long as $\lambda=0$). We find
\begin{equation}
 T_{\mu\nu} = (1-2\xi)\partial_\mu\varphi\partial_\nu\varphi -2 \xi \varphi \nabla_\mu\partial_\nu \varphi +2 \Bigl(\xi-\frac{1}{4}\Bigr) g_{\mu\nu} \partial_\rho\varphi\partial^\rho \varphi
 +2\xi g_{\mu\nu} \varphi \Box_g \varphi
 +\xi G_{\mu\nu} \varphi^2 -\frac{\lambda}{4!} g_{\mu\nu} \varphi^4
 \,,
\end{equation}
where $G_{\mu\nu}=R_{\mu\nu}-\frac{1}{2}R g_{\mu\nu}$ is the Einstein's tensor which comes from the variation of $R$ up to boundary terms.
On-shell we need the equations of motion
\begin{equation}
 \Box_g \varphi =\xi R \varphi +\frac{\lambda}{6} \varphi^3\,,
\end{equation}
and, substituting for $\Box_g\varphi$ solved from the equations of motion,
the trace $T^\mu{}_\mu$ becomes
\begin{equation}
 T^\mu{}_\mu = \Bigl(\xi-\frac{1}{6}\Bigr)\Bigl\{
 6 \partial_\rho\varphi\partial^\rho \varphi + 6 \xi R \varphi^2 + \lambda \varphi^4
 \Bigr\}\,,
\end{equation}
which, as expected, becomes zero for $\xi=\frac{1}{6}$, as implied by full Weyl invariance in the same limit. In agreement with the general analysis of harmonic invariance, we can verify that $T^\mu{}_\mu$ is the Laplacian of a scalar function
\begin{equation}
 T^\mu{}_\mu = \Box_g \Bigl\{ 3 \Bigl(\xi-\frac{1}{6}\Bigr) \varphi^2 \Bigr\}
 \,,
\end{equation}
where, to establish equality of the above two expressions, we have to use the equations of motion of $S[\varphi,g]$ once more. This confirms in practice that harmonic Weyl invariance implies that the trace of the energy-momentum tensor is the Laplacian of a scalar. Recall that simple scale invariance would only imply that $T^\mu{}_\mu$ is the divergence of the virial current \cite{Nakayama:2010zz,Zanusso:2023vkn}, so harmonic invariance is more constraining than scale invariance but less constraining than full Weyl invariance \cite{Edery:2014nha}.

The expression of the form $T^\mu{}_\mu= \Box_g \Phi$ gives the identification with the scalar operator introduced in \eqref{eq:harmonic-implication-T}
\begin{equation}
 \Phi[\varphi] = 3 \Bigl(\xi-\frac{1}{6}\Bigr) \varphi^2
 \,.
\end{equation}
Incidentally, this confirms that the original action is scale invariant by virtue of the fact that $T^\mu{}_\mu$ is the divergence of a vector $T^\mu{}_\mu= \nabla^\mu J_\mu$ on-shell, i.e., the virial current being $J_\mu = \partial_\mu \Phi$.

As mentioned in the previous part, the scalar operator can be coupled to $R$ and be used to ``improve'' the original action by adding the new interaction
\begin{equation}
\begin{split}
 S'[\varphi,g] &= S[\varphi,g] - \frac{1}{6} \int {\rm d}^4 x \sqrt{g} R \Phi[\varphi]
 \\
 &= S[\varphi,g]- \frac{1}{2} \int {\rm d}^4 x \sqrt{g} R \Bigl(\xi-\frac{1}{6}\Bigr) \varphi^2
\end{split}
\end{equation}
The different improved action $S'[\varphi,g]$ has now a traceless energy momentum tensor, so it is fully Weyl invariant. The net effect of the improvement is precisely to cancel out the $\xi$ dependence of $S[\varphi,g]$ as well as to replace it with the Weyl-invariant value
\begin{equation}
 S'[\varphi,g] = \int {\rm d}^4 x \sqrt{g}\Bigl\{\frac{1}{2}\partial_\mu\varphi\partial^\mu\varphi + \frac{1}{12} R \varphi^2 + \frac{\lambda}{4!}\varphi^4\Bigr\}\,.
\end{equation}
The improvement of the energy-momentum tensor of the harmonic Weyl invariant action $S[\varphi,g]$
comes at the price of losing the arbitrariness of coupling $\xi$.

\subsubsection{Quantum anomaly of harmonic Weyl invariance: a preliminary analysis}\label{subsubsect:anomaly-example-1}

Now we concentrate briefly on quantum anomalies.
Looking at \eqref{eq:harmonic-implication-T}, we expect an eventual quantum anomaly ${\cal A}$ of harmonic Weyl invariance to appear in the form
\begin{equation}
 \langle T^\mu{}_\mu \rangle = \Box_g \langle \Phi \rangle + {\cal A} + \bigl(\beta ~ {\rm terms}\bigr)\,,
\end{equation}
or, alternatively, ${\cal A}=\langle T^\mu{}_\mu - \Box_g \Phi\rangle $
at fixed points of the renormalization group flow for which the $\beta ~ {\rm terms}$ are zero.
Using standard covariant techniques based on the heat kernel method \cite{Vassilevich:2003xt}
and the background field method \cite{Abbott:1980hw}, we find for the action defined in Eq.~\eqref{dim4self} the one-loop contribution to the expectation value
of the trace of the EMT at the leading order
as
\begin{equation}
\begin{split}
(4\pi)^2\langle T^\mu{}_\mu \rangle|_{1-{\rm loop}}
=
&\frac{1}{180}\Bigl(
R^2_{\alpha\beta\mu\nu}
-
R^2_{\alpha\beta}
\Bigr)
+
\frac{1}{2}\Bigl(
\xi - \frac{1}{6}
\Bigr)^2 R^2
-
\frac{1}{6}\Bigl(
\xi - \frac{1}{6}
\Bigr) \Box_g R\\
&+
\frac{\lambda}{2}\Bigl(
\xi - \frac{1}{6}
\Bigr)  R \varphi^2
+
\frac{\lambda^2}{8}  \varphi^4
-
\frac{\lambda}{12} \Box_g  \varphi^2 \, ,
\end{split}
\end{equation}
where $\varphi$ in now the argument of the effective action $\Gamma[\varphi]$
(so it actually is $\langle \varphi\rangle$ and ours is a mild abuse of notation).
In this form, this is not the Weyl anomaly, because the original action was not Weyl invariant to begin with, neither it is straightforwardly interpreted as the
anomaly of \eqref{eq:harmonic-implication-T}. We need to write the result in a slightly different way first, because we need a better definition of the anomaly \cite{Ferrero:2023unz}.

The renormalization group fixed point of the beta function $\beta_\lambda=\frac{3}{(4\pi)^2}\lambda^2$ is $\lambda=0$. In this limit the scale dependence induced by the coupling $\lambda$ is eliminated and we can choose the correction
\begin{equation}
\begin{split}
(4\pi)^2\langle \Phi \rangle|_{1-{\rm loop}}
=
&-\frac{1}{6} \Bigl(
 \xi - \frac{1}{6}
 \Bigr) R \, ,
\end{split}
\end{equation}
so that, using $\langle T^\mu{}_\mu \rangle= T^\mu{}_\mu +\langle T^\mu{}_\mu \rangle|_{1-{\rm loop}}$ and $\langle \Phi \rangle = \Phi +\langle \Phi \rangle|_{1-{\rm loop}}$, their combination satisfies
\begin{equation}\label{eq:weyl-restricted-anomaly-simple-scalar}
\begin{split}
 (4\pi)^2 {\cal A} \equiv (4\pi)^2\langle T^\mu{}_\mu -\Box_g \Phi \rangle
 &=  \frac{1}{180}\Bigl(
 R^2_{\alpha\beta\mu\nu}
 -R^2_{\alpha\beta}
 \Bigr)
 +\frac{1}{2}\Bigl(
 \xi - \frac{1}{6}
 \Bigr)^2 R^2
\end{split}
\end{equation}
which gives us a meaningful form of the anomaly ${\cal A}$ of harmonic Weyl invariance. It is straightforward to rewrite ${\cal A}$ is a more natural basis that involves the square of the Weyl tensor and the four-dimensional Euler density
\begin{equation}\label{eq:weyl-restricted-anomaly-simple-scalar-WeylGB-basis}
\begin{split}
 (4\pi)^2 {\cal A} &=  \frac{1}{120}
 W^2_{\alpha\beta\mu\nu}
 -\frac{1}{360} E_4
 +\frac{1}{2}\Bigl(
 \xi - \frac{1}{6}
 \Bigr)^2 R^2\,,
\end{split}
\end{equation}
from which it becomes evident that, in the conformal limit, only the Weyl-integrable terms survive, in agreement with the standard cohomological analysis of the anomaly \cite{Bonora:1983ff}.
In addition to the general $a$ and $b$ anomalies \cite{Deser:1993yx}, in the general formula we also have a term proportional to $R^2$. Such a contribution, while prohibited by the Wess-Zumino integrability conditions of full Weyl invariance, is admissible for  harmonic Weyl invariance (see also appendix \ref{sect:appendix-cohomology}).

We believe that it is important to stress once more that the scalar $\Phi$ appearing in the classical identity \eqref{eq:T-improvement-R} has no ``operative'' definition, e.g., it is not the functional derivative of the action with respect to some source. This fact gives us the freedom to redefine it at one loop, and we used this
freedom to simplify the anomaly. In this way, we have obtained an anomaly that, in the conformal limit $\xi=\frac{1}{6}$ matches the standard trace anomaly of Weyl symmetry (see Refs.~\cite{Franchino-Vinas:2018gzr,Franchino-Vinas:2019upg} for a similar analysis that, instead, concentrates directly on the nonlocal part of $\Gamma$, which comes from integrating the anomaly). The term of the form $\Box_g R$, which has been cancelled by this choice, is technically also present in the Weyl anomaly, but in that case it is cohomologically trivial on the basis of the analysis of the BRST symmetry underlying standard Weyl invariance \cite{Bonora:1983ff}. Here, instead, we reabsorbed it in the redefinition of the scalar $\Phi$.

\subsubsection{Harmonic Weyl invariant scalar in $d=4$ with zero weight and its anomaly}\label{sect:restricted-example-2}

Now we switch our attention to a scalar field with zero weight, $w_\varphi=0$, so now $\varphi$ is invariant under Weyl transformations.
This case can be regarded as the $d\to 4$ limit of a higher derivative scalar with $w_\varphi=\frac{4-d}{2}$ ; the general $d$ case can also be considered in absence of self-interactions,
but we skip it for simplicity \cite{Edery:2014nha}.
We choose this example to highlight how harmonic Weyl invariance affects the freedom in the parameters of a general action, because the previous example of the weight one scalar enjoys a standard nonminimally coupled action, so it is not evident to what extent the harmonic symmetry may affect a more complicate model.
The most general harmonic Weyl invariant action with $\mathbb{Z}_2$ symmetry for a weight zero scalar with field-independent couplings in $d=4$ is
\begin{equation}
\begin{split}
 S[\varphi,g] = \int {\rm d}^4 x \sqrt{g}\Bigl\{&
 \frac{1}{2}\Box_g\varphi\Box_g\varphi
 -R^{\mu\nu} \partial_\mu \varphi \partial_\nu \varphi
 + \tilde{\xi} R \partial_\mu \varphi \partial^\mu \varphi
 + \tilde{\lambda} \partial_\mu \varphi \partial^\mu \varphi
 \partial_\nu \varphi \partial^\nu \varphi
 \Bigr\}\,,
\end{split}
\end{equation}
up to total derivatives and up to nonderivative terms of the form $\int\varphi^2 R$ and $\int \varphi^2 W^2$ that are manifestly invariant.
There are two simple ways to deduce the above action. On the one hand one can work by brute force and use an ansatz which includes all possible terms (including those that differ by total derivatives), then impose invariance under harmonic Weyl transformations.
On the other hand, we know that harmonic Weyl invariance constrains $R$ to transform homogeneously with weight two, so we can have an arbitrary coupling of $R$ with a weight two scalar constructed with $\varphi$ and the only option is $\partial_\mu\varphi\partial^\mu\varphi$. The quartic derivative interaction is generally Weyl invariant, at least classically, while the interaction with the Ricci tensor is fixed by harmonic Weyl invariance as if it was fully Weyl invariant. In fact, full conformal invariance is realized iff $\tilde{\xi}=\frac{1}{3}$, while $\tilde{\lambda}$ is always classically unconstrained. In terms of the number of couplings, this action mirrors closely the previous example, having one more than its conformal counterpart.

A rather lengthy, but straightforward, computation reveals a variational energy-momentum tensor $T_{\mu\nu}$, which is conserved on-shell as expected. The complete expression is rather long, but fortunately we only need the trace for our purposes.
We have that on-shell the trace becomes
\begin{equation}
\begin{split}
 T^\mu{}_\mu = 12 \Bigl(\tilde{\xi}-\frac{1}{3}\Bigr)\Bigl\{
 \Box_g \partial^\mu \varphi\partial_\mu \varphi + \nabla^\mu\partial^\nu \varphi\nabla_\mu\partial_\nu \varphi
 \Bigr\}\,.
\end{split}
\end{equation}
Using the equations of motion, it is easy to show that the above expression is equivalent to $T^\mu{}_\mu = \Box_g \Phi$
where
we deduce the scalar operator 
\begin{equation}
\begin{split}
 \Phi = 6 \Bigl(\tilde{\xi}-\frac{1}{3}\Bigr)\partial^\mu \varphi\partial_\mu \varphi\,.
\end{split}
\end{equation}
and the equations of motion must be used to prove the equality.
Much like in the previous example, we can ``improve'' the energy momentum tensor at the price of redefining the action $S' = S - \frac{1}{6}\int {\rm d}^4x \Phi R$  and the cost of this redefinition is precisely to cancel the $\tilde{\xi}$ coupling and replace it with the fully Weyl invariant value
\begin{equation}
\begin{split}
 S'[\varphi,g] = \int {\rm d}^4 x \sqrt{g}\Bigl\{&
 \frac{1}{2}\Box_g\varphi\Box_g\varphi
 -R^{\mu\nu} \partial_\mu \varphi \partial_\nu \varphi
 + \frac{1}{3} R \partial_\mu \varphi \partial^\mu \varphi
 + \tilde{\lambda} \partial_\mu \varphi \partial^\mu \varphi
 \partial_\nu \varphi \partial^\nu \varphi
 \Bigr\}\,.
\end{split}
\end{equation}
The propagation of the scalar field in the weak $\tilde{\lambda}$ limit is governed by the Paneitz-Riegert-Fradkin-Tseytlin Weyl-covariant operator, generally denoted $\Delta_4$ (that is, $\Delta_4$ is the Hessian of the above action in the limit $\tilde{\lambda}\to 0$) and discovered multiple times in the literature \cite{Fradkin:1982xc,paneitz,Riegert:1984kt}.
It is important to notice that this improvement procedure should be regarded as a classical one,
unless $\tilde{\lambda}=0$ or, as condidered later on, within the one loop approximation since the anomalous dimensions remain zero~\cite{Safari:2021ocb}, becoming non zero only at three loops. Indeed it has been shown in Ref.~\cite{Gimenez-Grau:2023lpz} that this shift-invariant theory is only scale invariant and not conformal at three loops in perturbation theory because the candidate virial current would also acquire a non trivial anomalous dimension.

Before moving on, notice that $S[\varphi, g]$ is not quite the most general action, because $\varphi$ is a dimensionless scalar, which opens the possibility for nonlinear field-dependent couplings, that are tensor functions of $\varphi$ itself (i.e., like a higher-derivative nonlinear sigma model \cite{Percacci:2009fh}). This example has been considered in the context of harmonic invariance in Ref.~\cite{Edery:2014nha}. The most general form of the Weyl invariant nonlinear sigma model action, which includes tensors with two, three and four indices in the target-space's ``coordinates'' $\varphi\to \varphi^a$ has previously appeared in Ref.~\cite{Romoli:2021hre} and the same approach can be used to derive the generalization of this section's example very easily. In particular, the four derivative action given in Ref.~\cite{Romoli:2021hre} is Weyl invariant without having to integrate by parts (so not up to boundary terms), which is particularly convenient when computing invariant actions and their generalizations.

As for the computation of the quantum anomaly, we follow closely the discussion given in Sect.~\ref{subsubsect:anomaly-example-1}. The computations of the leading $1$-loop divergence of the effective action
and of the expectation value of the EMT can be performed using once more covariant heat-kernel methods over an arbitrary background field. The two results are actually both based on the same heat kernel coefficient, that has been computed in full generality recently in Ref.~\cite{Barvinsky:2021ijq}, including, in particular, total derivatives, which are a necessity for our purpose. The coefficient gives essentially the leading contribution to the EMT
\begin{equation}\label{eq:T-expectation-value-example2}
\begin{split}
(4\pi)^2\langle T^\mu{}_\mu \rangle|_{1-{\rm loop}}
=
&\frac{1}{180}
R^2_{\alpha\beta\mu\nu}
-\frac{4}{45}
R^2_{\alpha\beta}
+
\Bigl(
\frac{5}{36}-\frac{2}{3}\tilde{\xi}+\tilde{\xi}^2
\Bigr) R^2
+\Bigl(
\frac{23}{90}-\frac{2}{3}\tilde{\xi}
\Bigr) \Box_g R\\
&
+6 \Bigl(\tilde{\xi}-\frac{1}{3}\Bigr) \tilde{\lambda} R (\partial\varphi)^2
+10 \tilde{\lambda}^2 \partial_\mu \varphi \partial^\mu \varphi \partial_\nu\varphi \partial^\nu \varphi
\\
&
-\frac{8}{9} \tilde{\lambda} (\Box_g\varphi)^2
-\frac{16}{3} \tilde{\lambda}\partial_\mu \Box_g\varphi \partial^\mu \varphi
-\frac{40}{9} \tilde{\lambda}\nabla_\mu \partial_\nu \varphi \nabla^\mu\partial^\nu \varphi
-\frac{40}{9}\tilde{\lambda} R^{\mu\nu} \partial_\mu\varphi\partial_\nu\varphi
\, ,
\end{split}
\end{equation}
which includes several terms that, in principle, are allowed by the cohomological analysis (see appendix~\ref{sect:appendix-cohomology}) that we address in a moment.
The one-loop beta function of
$\tilde{\lambda}$ can be found from the leading divergent part of the effective action in dimensional regularization,
which also coincides with the integrated form of the above expression up to a factor $2$ and the $\epsilon$-pole, so we find that
\begin{equation}
\beta_{\tilde{\lambda}}=\frac{5}{(4\pi)^2}\tilde{\lambda}^2\,,
\end{equation}
but also that there is no anomalous dimension for the $\varphi$ field (the integration by parts of the terms with four derivatives gives zero) in agreement with Refs.~\cite{Safari:2021ocb,Tseytlin:2022flu} (see also Refs.~\cite{Buccio:2022egr,Buccio:2023lzo,Holdom:2023usn} for applications in the context of physical amplitudes and unitarity).

To handle the additional terms, we rewrite \eqref{eq:T-expectation-value-example2} as
\begin{equation}\label{eq:T-expectation-value-example2-step2}
\begin{split}
 (4\pi)^2\langle T^\mu{}_\mu \rangle|_{1-{\rm loop}}
 =
 &\frac{1}{180}
 R^2_{\alpha\beta\mu\nu}
 -\frac{4}{45}
 R^2_{\alpha\beta}
 +
 \Bigl(
  \frac{5}{36}-\frac{2}{3}\tilde{\xi}+\tilde{\xi}^2
 \Bigr) R^2
 +\Bigl(
  \frac{23}{90}-\frac{2}{3}\tilde{\xi}
 \Bigr) \Box_g R
 \\
 &
 +6 \Bigl(\tilde{\xi}-\frac{1}{3}\Bigr) \tilde{\lambda} R (\partial\varphi)^2
 +10 \tilde{\lambda}^2 \partial_\mu \varphi \partial^\mu \varphi \partial_\nu\varphi \partial^\nu \varphi
 +\nabla_\mu V^\mu
 \, ,
\end{split}
\end{equation}
where we introduced the vector $V_\mu=-\frac{8}{9}\tilde{\lambda}\bigl(\partial_\mu\varphi \Box_g\varphi +\frac{5}{2} \nabla_\mu(\partial\varphi)^2\bigr)$, which gives a ``trivial'' anomaly of full Weyl symmetry. The trivial anomaly $\nabla_\mu V^\mu$, which corresponds to the entire third line of \eqref{eq:T-expectation-value-example2}, can be eliminated by an opportune choice of the renormalization scheme of the $\varphi$ Lagrangian.

At this point notice that, with the elimination of $\nabla_\mu V^\mu$, all the remaining terms in \eqref{eq:T-expectation-value-example2-step2} are compatible with the invariance under harmonic Weyl transformations. At the fixed point $\tilde{\lambda}=0$ of $\beta_{\tilde{\lambda}}=0$, we can define
\begin{equation}
\begin{split}
(4\pi)^2\langle \Phi \rangle|_{1-{\rm loop}}
=
&\Bigl(
\frac{23}{90}-\frac{2}{3}\tilde{\xi}
\Bigr) R \, ,
\end{split}
\end{equation}
that modifies the classical operator $\Phi$ by a quantum correction proportional to the curvature, as done in Sect.~\ref{subsubsect:anomaly-example-1}.
Given the crafted definition of $\langle\Phi\rangle=\Phi + \langle \Phi \rangle|_{1-{\rm loop}}$, we find the anomaly of harmonic Weyl symmetry at the fixed point as
\begin{equation}\label{eq:weyl-restricted-anomaly-hd-scalar}
\begin{split}
 (4\pi)^2 {\cal A} \equiv (4\pi)^2\langle T^\mu{}_\mu -\Box_g \Phi \rangle
 &=  \frac{1}{180}
R^2_{\alpha\beta\mu\nu}
-\frac{4}{45}
R^2_{\alpha\beta}
+
\Bigl(
\frac{5}{36}-\frac{2}{3}\tilde{\xi}+\tilde{\xi}^2
\Bigr) R^2\,,
\end{split}
\end{equation}
and includes the same geometric structures as the previous example. We have second checked most of this result using the heat-kernel coefficients given in the classic Ref.~\cite{Barth:1985sy}, which gives them up to the boundary terms (so up to our definition of $\Box_g \Phi$) given in Ref.~\cite{Barvinsky:2021ijq}.

Finally, we write the anomaly in terms of the natural basis that includes the square of the Weyl tensor and the four-dimensional Euler density
\begin{equation}\label{eq:weyl-restricted-anomaly-hd-scalar-integrable}
\begin{split}
 (4\pi)^2 {\cal A} \equiv (4\pi)^2\langle T^\mu{}_\mu -\Box_g \Phi \rangle
 &=  -\frac{1}{30}
 W^2_{\alpha\beta\mu\nu}
 +\frac{7}{180}
 E_4
 +
 \Bigl(
 \tilde{\xi}-\frac{1}{3}
 \Bigr)^2 R^2\,,
\end{split}
\end{equation}
which gives, as expected, only the fully Weyl integrable terms in the conformal limit $\tilde{\xi}=\frac{1}{3}$ \cite{Riegert:1984kt}.
Amusingly the last term that is not allowed by integrability
has a ``charge'' that is proportional to the square of the deviation of the coupling $\tilde{\xi}$ from it conformal value. This situation mimicks precisely the example of the scalar field of dimension one.

\subsection{Implications of Liouville-Weyl invariance}
\label{sect:liouville-implications}

Now we turn our attention to the Liouville-Weyl groupoid discussed originally in Sect.~\ref{sect:liouville-group}. We follow the same strategy as in the harmonic case of Sect.~\ref{sect:harmonic-implications}, but generalize it appropriately to a differential constraint with two uncontracted indices.

We begin by considering the infinitesimal form of the Liouville-Weyl transformations
We can and argue that a theory with action $S[\Psi]$ that is Liouville-Weyl invariant, but not generally Weyl invariant, must be such that on-shell
\begin{equation}\label{eq:liouville-implication-mistake}
 \int {\rm d}^dx \sqrt{g} \bigl(
 T^\mu{}_\mu \sigma
 - X^{\mu\nu} \, \nabla_\mu \partial_\nu \sigma |_{\rm traceless}
 \bigr)=0\,,
\end{equation}
where $\nabla_\mu \partial_\nu \sigma |_{\rm traceless}= \nabla_\mu \partial_\nu \sigma -\frac{1}{d}g_{\mu\nu} \Box_g\sigma$ and, for consistency, $X_{\mu\nu}$ must be a traceless tensor constructed with the field.
As for the harmonic case we can consider a finite transformation. Using results and notation of Sect.~\ref{sect:liouville-group}, we know that it is possible to write the finite transformations in a way that resembles the infinitesimal ones. Therefore, we can make the replacements
\begin{equation}\label{eq:liouville-implication-correct}
\begin{split}
 \delta_\sigma S[\Psi,g] \to 
 S[\Omega^{w_\Psi}\Psi,\Omega^2 g]-S[\Psi,g]\,,
 \qquad \qquad
 \nabla_\mu \partial_\nu \sigma|_{\rm traceless} \to \Omega^2 P_{g,\mu\nu} (\Omega^{-1})=0
 \,,
\end{split}
\end{equation}
which includes variations beyond the linear order.
Using \eqref{eq:liouville-implication-mistake},
we see that the requirement $\delta_\sigma S[\Psi,g]=0$ implies in general
\begin{equation}\label{eq:emt-liouville-implication}
 T^\mu{}_\mu= \nabla_\mu \nabla_\nu X^{\mu\nu}\,,
\end{equation}
so that the trace of the EMT is the double divergence of a symmetric traceless tensor.
Exactly like in the harmonic case, this condition on the trace of the EMT is stronger than simple
scale invariance, because this $T_{\mu\nu}$ can always be ``improved'' to a traceless form by adding a new coupling to $S$. In flat space the metric is $g_{\mu\nu}=\eta_{\mu\nu}$ and the improvement would require
\begin{equation}
 T^{\mu\nu} \to T'^{\mu\nu}=T^{\mu\nu} + {\cal D}^{\mu\nu}{}_{\rho\theta}  X^{\rho\theta}
\end{equation}
where we defined the flat space's differential operator
\begin{equation}
 {\cal D}^{\mu\nu\rho\theta} =
 \frac{1}{d-2}(\eta^{\mu(\rho} \partial^{\theta)} \partial^{\nu}
 +\eta^{\nu(\rho} \partial^{\theta)} \partial^{\mu}
 -\eta^{\mu(\rho} \eta^{\theta)\nu} \partial^2
 -\eta^{\mu\nu}\partial^\rho\partial^\theta)
 -\frac{1}{(d-1)(d-2)}\eta^{\rho\theta}(\partial^\mu\partial^\nu -\eta^{\mu\nu}\partial^2)
\end{equation}
as in Refs.~\cite{Stergiou:2022qqj}.
With a bit of work one can show that $T'{}^\mu{}_\mu=0$ and that this procedure
is completely unique in $2<d<6$, while in $d\geq 6$ additional structures related to the Bach tensor may be present \cite{Osborn:2016bev,Osborn:2015rna,Stergiou:2022qqj} and are required to construct a $T'^{\mu\nu}$ that is also a primary operator of the flat space's CFT. In order to interpret the new $T'^{\mu\nu}$
as (the flat space limit of) a variational EMT, the original action
must be modified by including a new coupling between $X_{\mu\nu}$ and the Schouten tensor defined in \eqref{eq:schouten-def}. This is done as $S[\Psi,g] \to S'[\Psi,g]=S[\Psi,g]+\Delta S[\Psi,g]$,
where
\begin{equation}
 \Delta S[\Psi,g] = - \int {\rm d}^d x \sqrt{g} K_{\mu\nu} X^{\mu\nu}\,.
\end{equation}
Notice that this strategy of improvement is very similar to the one given in Sect.~\ref{sect:harmonic-implications}, because, in that case, the improvement can be seen as a coupling of $\Phi$ with the trace of the Schouten tensor ${\cal J}$ defined in \eqref{eq:J-def} (see also Sect.~\ref{sect:general-recipe}). However, in the case of Liouville-Weyl transformations, we have that $X_{\mu\nu}$ is traceless, so only the traceless part of the Schouten tensor $\tilde{K}_{\mu\nu} \equiv K_{\mu\nu}-\frac{1}{d}{\cal J}g_{\mu\nu}$ is actually coupled to $X_{\mu\nu}$. Given that, up to a dimensionality dependent factor the tensors $K_{\mu\nu}$ and $R_{\mu\nu}$ differ only by their trace, the same coupling can be seen as a coupling with the traceless part of the Ricci tensor, 
$\tilde{R}_{\mu\nu} \equiv R_{\mu\nu}-\frac{1}{d}R g_{\mu\nu}$.

\subsubsection{Example: Liouville-Weyl invariant weight zero scalar in $d=4$ and its anomaly}

Any two-derivative action of a scalar with weight $w_\varphi=1$ in $d=4$
that is Liouville-Weyl invariant is also generally Weyl invariant. This happens because there is no way to couple a simple weight one scalar to a symmetric traceless
tensor and produce a weight zero density (we are restricting our attention to perturbatively renormalizable terms), so we do not have an example for Liouville-Weyl which is analog to Sect.~\ref{sect:restricted-example-1}. Things change if the field is allowed to have indices, as we later discuss in Sect.~\ref{sect:general-recipe}.

We do have an example analog to Sect.~\ref{sect:restricted-example-2} with a weight zero scalar, however.
The most general Liouville-Weyl invariant action for a scalar with $w_\varphi=0$ is
\begin{equation}
\begin{split}
 S[\varphi,g] = \int {\rm d}^4 x \sqrt{g}\Bigl\{&
 \frac{1}{2}\Box_g\varphi\Box_g\varphi
 -\frac{1}{6} R \partial_\mu \varphi \partial^\mu \varphi
 + \tilde{\zeta} \tilde{R}^{\mu\nu} \partial_\mu \varphi \partial_\nu \varphi
 + \tilde{\lambda} \partial_\mu \varphi \partial^\mu \varphi
 \partial_\nu \varphi \partial^\nu \varphi
 \Bigr\}
\end{split}
\end{equation}
up to boundary terms, and, as before, $\tilde{R}_{\mu\nu}$ is the traceless Ricci tensor.
As expected, we now have a dimension-two tensor $\partial_\mu\varphi\partial_\nu\varphi$, which couples to the (traceless part) of the Ricci tensor.
A lengthy, but straightforward, computation reveals a variational energy-momentum tensor $T_{\mu\nu}$, which, on-shell, has trace
\begin{equation}
\begin{split}
 T^\mu{}_\mu = \nabla_\mu \nabla_\nu X^{\mu\nu}\,.
\end{split}
\end{equation}
This is the expected form as in \eqref{eq:emt-liouville-implication}, with the symmetric traceless tensor $X_{\mu\nu}$ being 
\begin{equation}
\begin{split}
 X_{\mu\nu} &= 2 (\tilde{\zeta}+2) \Bigl\{
 \partial_\mu \varphi \partial_\nu \varphi -\frac{1}{4} g_{\mu\nu} \partial_\rho\varphi \partial^\rho\varphi
 \Bigr\} \,.
\end{split}
\end{equation}
The equations of motion must be used to prove the above expressions, which are only true on-shell.

In analogy to all previous examples, at the price of Liouville-Weyl invariance, we could ``improve'' the energy momentum tensor to a traceless form.
This requires the change of  the action to a new one, $S' = S -\int {\rm d}^4x K^{\mu\nu} X_{\mu\nu}$ and the cost of this redefinition is precisely the $\tilde{\zeta}$ coupling, which is replaced with its fully Weyl invariant value.\footnote{Nevertheless, it has been recently observed in Ref.~\cite{Gimenez-Grau:2023lpz} that, in flat space and at the interacting nontrivial RG fixed point, the shift-invariant theory is only scale invariant and not conformal at three loops in perturbation theory because the anomalous dimension becomes nonzero. Therefore, the improvement that we discuss here is only classical, unless $\tilde{\lambda}=0$ or the computation is performed in the one loop approximation.} Ultimately the improved $S'$ is the same as in Sect.~\ref{sect:restricted-example-2}.

Now we concentrate on the quantum anomaly of Liouville-Weyl invariance, following an analog strategy to Sect.~\ref{sect:restricted-example-2}.
Similarly to the harmonic case, we first look at \eqref{eq:emt-liouville-implication}, and deduce that an eventual quantum anomaly ${\cal A}$ of Liouville-Weyl invariance should be of the form
\begin{equation}
 \langle T^\mu{}_\mu \rangle = \nabla_\mu \nabla_\nu \langle X^{\mu\nu} \rangle + {\cal A} + \bigl(\beta ~ {\rm terms}\bigr)\,,
\end{equation}
or, alternatively, ${\cal A}=\langle T^\mu{}_\mu - \nabla_\mu \nabla_\nu X^{\mu\nu}\rangle $
at fixed points of the renormalization group flow for which the $\beta ~ {\rm terms}$ are zero (in this section's example, $\beta_{\tilde{\lambda}}=0$ ). 
Using the heat kernel coefficients given in Ref.~\cite{Barvinsky:2021ijq},
and the background field method, we find the one-loop contribution to the expectation value
of the trace of the EMT at the leading order
as
\begin{equation}
\begin{split}
(4\pi)^2\langle T^\mu{}_\mu \rangle|_{1-{\rm loop}}
=
&\frac{1}{180}
R^2_{\alpha\beta\mu\nu}
+\frac{1}{180}(44+60\tilde{\zeta}+15\tilde{\zeta}^2)
R^2_{\alpha\beta}
-\frac{1}{144}(8+12\tilde{\zeta}+3\tilde{\zeta}^2) R^2
\\&
-
\frac{1}{90}(7+5\tilde{\zeta})\Box_g R
+\frac{2}{3}(2+\tilde{\zeta}) \tilde{\lambda}\Bigl(
R_{\mu\nu}-\frac{1}{4}g_{\mu\nu}R
\Bigr) \partial^\mu \varphi\partial^\nu \varphi
\\
&
+ 10 \tilde{\lambda}^2 \partial_\mu \varphi \partial^\mu\varphi \partial_\nu \varphi \partial^\nu\varphi
+\nabla_\mu V^\mu
\end{split}
\end{equation}
where $\varphi$ in now the argument of the effective action $\Gamma[\varphi]$
as in the previous case. For brevity, we have also already separated the trivial anomaly $\nabla_\mu V^\mu$, which coincides with the same vector given before in Sect.~\ref{sect:restricted-example-2}. The interaction term contains the beta function of $\tilde{\lambda}$, which also coincides with Sect.~\ref{sect:restricted-example-2}.

At the renormalization group fixed point $\tilde{\lambda}=0$, we choose the correction
\begin{equation}
\begin{split}
(4\pi)^2\langle X_{\mu\nu} \rangle|_{1-{\rm loop}}
 =
 & \frac{2}{45}(7+5 \tilde{\zeta}) \tilde{R}_{\mu\nu}
 =
 \frac{2}{45}(7+5 \tilde{\zeta}) \Bigl(R_{\mu\nu}-\frac{1}{4}R g_{\mu\nu}\Bigr)
 \, ,
\end{split}
\end{equation}
that respects the fact that any correction to $\langle X_{\mu\nu}\rangle$ must be traceless.
Once more, we remark that the choice is possible because there is no operative definition of the (renormalized) tensor $X_{\mu\nu}$, and this is the same freedom that we discussed and exploited in Sects.~\ref{sect:restricted-example-1}, \ref{subsubsect:anomaly-example-1} and \ref{sect:restricted-example-2} for $\Phi$, except for the tracelessness condition.
Then, using $\langle T^\mu{}_\mu \rangle= T^\mu{}_\mu +\langle T^\mu{}_\mu \rangle|_{1-{\rm loop}}$ and $\langle X_{\mu\nu} \rangle = X_{\mu\nu} +\langle X_{\mu\nu} \rangle|_{1-{\rm loop}}$, we find that their combination satisfies
\begin{equation}\label{eq:weyl-liouville-anomaly-simple-scalar}
\begin{split}
 (4\pi)^2 {\cal A} &\equiv (4\pi)^2\langle T^\mu{}_\mu - \nabla_\mu \nabla_\nu X^{\mu\nu} \rangle
 \\&
 =  \frac{1}{180}
R^2_{\alpha\beta\mu\nu}
+\frac{1}{180}(44+60\tilde{\zeta}+15\tilde{\zeta}^2)
R^2_{\alpha\beta}
-\frac{1}{144}(8+12\tilde{\zeta}+3\tilde{\zeta}^2) R^2
\,,
\end{split}
\end{equation}
which gives us a form of the anomaly ${\cal A}$ of Liouville-Weyl invariance. If we choose to rewrite ${\cal A}$ in the integrable basis that involves the square of the Weyl tensor and the four-dimensional Euler density, we find
\begin{equation}\label{eq:weyl-liouville-anomaly-simple-scalar-WeylGB-basis}
\begin{split}
 (4\pi)^2 {\cal A} &=  \frac{1}{120}(16+20\tilde{\zeta}+5\tilde{\zeta}^2)
 W^2_{\alpha\beta\mu\nu}
 -\frac{1}{360} (46+60\tilde{\zeta}+15\tilde{\zeta}^2) E_4
+\frac{1}{144} (2+\tilde{\zeta})^2 R^2\,,
\end{split}
\end{equation}
from which we also see that, in the conformal limit $\tilde{\zeta}=-2$, only the Weyl-integrable terms survive as expected by the cohomological analysis \cite{Bonora:1983ff}. We can go one step further and adopt the basis in which $R^2$ is replaced by $\tilde{R}_{\mu\nu}^2$, which is a more natural scalar invariant in the case of Liouville-Weyl invariance, in which case we get
\begin{equation}\label{eq:weyl-liouville-anomaly-simple-scalar-WeylGB-basis-2}
\begin{split}
(4\pi)^2 {\cal A} &= -\frac{1}{30} W^2_{\alpha\beta\mu\nu} +\frac{7}{180}E_4
 +\frac{1}{12} (2+\tilde{\zeta})^2 \tilde{R}_{\mu\nu}^2\,.
\end{split}
\end{equation}
Quite amusingly
the form \eqref{eq:weyl-liouville-anomaly-simple-scalar-WeylGB-basis-2} is fully analogous to
\eqref{eq:weyl-restricted-anomaly-hd-scalar-integrable}, after having chosen a basis which includes the square of the tensor that is left invariant by the symmetry (either $R$ or $\tilde{R}_{\mu\nu}$ 
 for the Harmonic or the Liouville-Weyl cases, respectively). In both cases, and also analogously to the case of Eq.~\eqref{eq:weyl-restricted-anomaly-simple-scalar-WeylGB-basis}, the last term has 
the coefficient of the new curvature square invariant term multiplied by a polynomial with a degenerate (double) root at the conformal value.

\section{General recipe for constructing groupoids invariant actions}
\label{sect:general-recipe}

In this section we outline a general strategy to derive two-derivative quadratic actions for irreducible tensors that are invariant under either the harmonic or the Liouville-Weyl subgroups. In order to do this we exploit a similar strategy to that used in Ref.~\cite{Erdmenger:1997wy} for finding the conformal action of completely symmetric traceless tensors.

To fix ideas, let us consider a matter field tensor $\Psi^{\mu_1 \dots \mu_p}{}_{\nu_1 \dots \nu_q}$ that lives on some irreducible tensor representation of $GL(d)$. 
We further require $\Psi$ to be traceless, because any trace, e.g., $\Psi^{\mu_1 \dots \mu_p}{}_{\mu_1 \dots \nu_q}$, can be treated as a tensor in a different representation with less indices.
Finally, the tensor is also required to transform homogeneously under Weyl transformations with a weight $w$ that is fixed to be $w=\frac{2q-2p+2-d}{2}$ by requiring invariance under scale transformations and covariant terms with up to two derivatives. The weight can be naturally inferred by counting how many copies of $g^{\mu\nu}$ are required to construct scale invariant kinetic terms.

The general quadratic action for $\Psi$ can be written as
\begin{equation}
 S[\Psi,g] = S_0[\Psi,g] + S_1[\Psi,g] \, ,
\end{equation}
where $S_0$ contains the sum of all possible kinetic terms (contractions of two copies of $\Psi$ and two derivatives), which are guaranteed to be scale invariant by our weight choice, while $S_1$ depends linearly on curvature tensors. As in the method of Ref.~\cite{Erdmenger:1997wy}, these curvatures will be linear in the Schouten tensor and adjusted to balance the transformation of $S_0$. Since all the nontrivial contractions of $\Psi$ to the Weyl tensor are trivially Weyl invariant, we do not need to take them into account here,
but rather we add them later through new interactions of the form
$S_3 \sim \int \sqrt{g} \Psi^2 W$ and appropriate contractions.

Given that $S_0$ is scale invariant, it must be possible to write its infinitesimal Weyl transformation as the contraction of a vector and the derivative of the scale factor
\begin{equation}
 \delta_\sigma S_0[\Psi,g] = \int {\rm d}^d x \sqrt{g} \, j^\lambda \partial_\lambda \sigma \, ,
\end{equation}
so that we would have $\delta_\sigma S_0=0$ for constant $\sigma$ \cite{Erdmenger:1997wy}.
Now notice that $\delta_\sigma K_{\mu\nu}$ depends quadratically on the derivatives of $\sigma$ and we want to use $K_{\mu\nu}$ to balance the transformation of $S_0$. This is possible iff $j^\lambda$ can be expressed as the divergence of a symmetric tensor, i.e.,
\begin{equation}\label{eq:Erdmenger-trick}
 j^\lambda = - \nabla_\rho J^{\rho\lambda} \, .
\end{equation}
This requirement uniquely fixes all the relative coefficients of the terms that involve divergences of $\Psi$ in terms of normalization constant of the minimal term of the kinetic action if full Weyl invariance is required through the coupling between $J_{\mu\nu}$ and $K_{\mu\nu}$. See, e.g., Refs.~\cite{Erdmenger:1997gy,Erdmenger:1997wy,Paci:2023twc}
for several examples.

Invariance under the harmonic and Liouville-Weyl subgroups of the Weyl group, instead, can be studied by splitting $J^{\rho\lambda}$ into its traceless and trace parts, that is,
\begin{align}\label{eq:split-Erdmenger-2tensor}
 J^{\lambda\rho} = X_0^{\lambda\rho} + g^{\lambda\rho} \Phi_0 \, ,
\end{align}
where
\begin{eqnarray}\label{eq:split-Erdmenger-2tensor-relations}
 X_0^{\lambda\rho} = J^{\lambda\rho} - \frac{1}{d} g^{\lambda\rho} J^\mu{}_\mu \, , && \qquad \Phi_0 = \frac{1}{d} J^\mu{}_\mu \, .
\end{eqnarray}
Given that $\Psi$ is an irreducible tensor, there is only one scalar contraction of two $\Psi$s. On the other hand, the number of tensor structures that enters the explicit expression of $X_0^{\lambda\rho}$ depends on the specific representation of $\Psi$ and can easily exceed one.

The requirement that $\Psi$ is traceless implies that $J^{\lambda\rho}$ is the sum of all possible symmetric two-tensors of the form $\Psi^{\mu\mu_2 \dots \mu_p}{}_{\nu_1 \dots \nu_q} \Psi^\nu{}_{\mu_2 \dots \mu_p}{}^{\nu_1 \dots \nu_q}$, i.e., the two uncontracted indices cannot come from the same $\Psi$. For example, for differential forms and completely symmetric traceless tensors we only have one independent such symmetric traceless two-tensor \cite{Erdmenger:1997gy,Erdmenger:1997wy}, while for mixed-symmetry tensors there can be more than one, i.e., we can build two nontrivial such tensors out of traceless hook-symmetric and hook-antisymmetric tensors \cite{Paci:2023twc}. 

We begin with the analysis of the harmonic Weyl invariance.
Using the decomposition given in Eq.~\eqref{eq:split-Erdmenger-2tensor} in Eq.~\eqref{eq:Erdmenger-trick}, we see that the infinitesimal transformation of the kinetic action now reads
\begin{equation}
 \delta_\sigma S_0[\Psi,g] = \int {\rm d}^d x \sqrt{g} \Bigl\{ X^{\mu\nu}_0  \Bigl( \nabla_\mu \partial_\nu - \frac{1}{d} g_{\mu\nu} \Box_g \Bigr) \sigma + \Phi_0 \Box_g \sigma \Bigr\} \, .
\end{equation}
We find convenient to use the traceless part of the Schouten tensor
\begin{equation}
 \tilde{K}_{\mu\nu} \equiv K_{\mu\nu} - \frac{1}{d} g_{\mu\nu} \mathcal{J} \, ,
\end{equation}
whose infinitesimal Weyl transformation is given by
\begin{equation}
 \delta_\sigma \tilde{K}_{\mu\nu} = \Bigl( \nabla_\mu \partial_\nu - \frac{1}{d} g_{\mu\nu} \Box_g \Bigr) \sigma \, .
\end{equation}
Therefore, it is straightforward to write a harmonic Weyl invariant action as
\begin{equation}\label{eq:gen-har-inv-action}
 S_{\rm har} [\Psi,g] = S_0[g,\Psi] + \int {\rm d}^dx \sqrt{g} \Bigl(  X_0^{\mu\nu} \tilde{K}_{\mu\nu} + \alpha_1 \Psi^2 \mathcal{J} \Bigr) \,,
\end{equation}
with $\alpha_1$ an arbitrary coupling and
up to eventual terms that depend on contractions with the Weyl tensor that are also fully Weyl invariant. Thus, two-derivative harmonic Weyl invariant quadratic actions for traceless $SO(d)$-irrep tensors depend on one more free parameter, i.e., $\alpha_1$, if compared to fully conformal actions for the same tensors.

Now we switch to the analysis of Liouville-Weyl invariance.
To begin, let us define the symmetric tensor $M^{\mu\nu}$ as
\begin{align}
 M^{\mu\nu} \equiv \sum_i a_i \Psi^{\dots \mu \dots}{}_{\dots} \Psi_{\dots}{}^\nu{}_{\dots}{}^{\dots} + \sum_j b_j \Psi^{\dots}{}_{\dots}{}^\mu{}_{\dots} \Psi_{\dots}{}^{\dots \nu \dots} + \sum_k c_k \Psi^{\dots \mu \dots}{}_{\dots} \Psi_{\dots}{}^{\dots\nu\dots} \, ,
\end{align}
where the sums cover all the inequivalent ways of choosing the position of the indices $\mu$ and $\nu$, while contracting all the other indices. The $a_i$, $b_j$ and $c_k$ are some coupling constants. We then denote by $\tilde{M}^{\mu\nu}$ its traceless part, i.e.,
\begin{equation}
 \tilde{M}^{\mu\nu} \equiv M^{\mu\nu} - \frac{1}{d} g^{\mu\nu} M^\lambda{}_\lambda\,.
\end{equation}
It is then straightforward to write the most general Liouville-Weyl invariant action as
\begin{equation}\label{eq:gen-liov-inv-action}
 S_{\rm LW} [\Psi,g] = S_0[g,\Psi] + \int {\rm d}^dx \sqrt{g} \Bigl(  \tilde{M}^{\mu\nu} \tilde{K}_{\mu\nu} + \Phi_0 \mathcal{J} \Bigr) \, ,
\end{equation}
up to terms that involve fully Weyl invariant constractions with the Weyl tensor.
The requirement of Liouville-Weyl invariance does not fix the couplings appearing in the parametrization of $\tilde{M}_{\mu\nu}$. Hopefully, the procedure is further clarified by two examples below.

\subsection{Example: groupoids invariant vector one-form actions}

Consider the most general two-derivative action that is quadratic in a one-form $A_\mu$
\begin{equation}
 S_0 [A,g] = \frac{1}{2} \int {\rm d}^dx \sqrt{g} \Bigl\{ (\nabla_\mu A_\nu) (\nabla^\mu A^\nu) + \gamma (\nabla_\mu A^\mu)^2  \Bigr\}\, .
\end{equation}
The requirement of scale invariance fixes the Weyl weight $w=\frac{4-d}{2}$
for the field $A_\mu$. Following Refs.~\cite{Erdmenger:1997gy,Paci:2023twc},
one first shows that $j^\mu$ can be written as the divergence of a symmetric tensor iff $\gamma=-\frac{4}{d}$. Similarly,
we can read the explicit form of the symmetric tensor $J^{\lambda\rho}$
\begin{equation}
 J^{\lambda\rho} = A^\rho A^\lambda + \frac{d-2}{2} A^2 g^{\rho\lambda} \, .
\end{equation}
Using the notation of Eq.~\eqref{eq:split-Erdmenger-2tensor-relations}, we find the traceless and trace parts
\begin{eqnarray}\label{eq:Erd-split-vector-case}
 X_0^{\rho\lambda} = A^\rho A^\lambda - \frac{1}{d} A^2 g^{\rho\lambda} \, , \qquad && \Phi_0 = \frac{d^2-2d+2}{2d} A^2 \, .
\end{eqnarray}
Now, knowing the tensor $\Phi_0$ and using Eq.~\eqref{eq:gen-har-inv-action}, we deduce the most general harmonic invariant action that is quadratic in the vector $A_\mu$, i.e.,
\begin{equation}\label{eq:harmonic-vector}
 S_{\rm har} [A,g] = \frac{1}{2} \int {\rm d}^dx \sqrt{g} \Bigl\{ (\nabla_\mu A_\nu) (\nabla^\mu A^\nu) - \frac{4}{d} (\nabla_\mu A^\mu)^2 + 2 \Bigl(A^\mu A^\nu - \frac{1}{d} g^{\mu\nu} A^2 \Bigr) K_{\mu\nu} + \Upsilon_1 \mathcal{J} A^2  \Bigr\} \, ,
\end{equation}
and it depends on one arbitrary coupling $\Upsilon_1$.
Analogously, the general Liouville-Weyl invariant action for the one-form is found by coupling $X_0^{\rho\lambda}$ of \eqref{eq:Erd-split-vector-case} in Eq.~ \eqref{eq:gen-liov-inv-action}
\begin{align}\label{eq:liouville-vector}
  S_{\rm LW} [A,g] = \frac{1}{2} \int {\rm d}^dx \sqrt{g} & \Bigl\{ (\nabla_\mu A_\nu) (\nabla^\mu A^\nu) - \frac{4}{d} (\nabla_\mu A^\mu)^2 + \frac{d^2-2d+2}{d} \mathcal{J} A^2\\\nonumber
  & + \Upsilon_2 \Bigl(A^\mu A^\nu - \frac{1}{d} g^{\mu\nu} A^2 \Bigr) K_{\mu\nu}  \Bigr\} \, .
\end{align}
Thus, according to the general analysis, we have that both the harmonic and Lioville subgroups introduce a free parameter in the invariant action of a vector, while the conformal counterpart is fully determined.

Given the above actions, we ask the question: what type of dynamical field is $A_\mu$?
In $d=4$, full Weyl invariance is compatible with gauge $U(1)$ symmetry $A_\mu \to A_\mu +\partial_\mu \Lambda$. To see this, we first derive the fully Weyl-invariant action in general $d$. To save time, notice that it can be obtained either inserting $\Upsilon_2=2$ in \eqref{eq:harmonic-vector}, or alternatively inserting $\Upsilon_1=\frac{d^2-2d+2}{d}$ in \eqref{eq:liouville-vector}. After an integration by parts, the fully Weyl invariant action becomes
\begin{align}\label{eq:weyl-vector-d4}
  S_{\rm Weyl} [A,g] = \int {\rm d}^dx \sqrt{g} & \Bigl\{
  \frac{1}{4}  F_{\mu\nu} F^{\mu\nu} + \frac{d-4}{2d} (\nabla_\mu A^\mu)^2
  - \frac{d-4}{2} K_{\mu\nu} A^\mu A^\nu + \frac{d-4}{2} {\cal J} A^\mu A_\mu  \Bigr\} \, ,
\end{align}
where $F_{\mu\nu}=\partial_\mu A_\nu-\partial_\nu A_\mu$ is the curvature two-form of $A_\mu$. The ``emergence'' of gauge $U(1)$ invariance in the limit $d=4$ should be evident by the fact that only the coupling of the $U(1)$-invariant curvature $F_{\mu\nu}$
survives and the other interactions decouple \cite{Erdmenger:1997gy}. So, in $S_{\rm Weyl} [A,g]$
we have that $A_\mu$ is a $U(1)$ gauge field iff $d=4$ \cite{El-Showk:2011xbs}.

In contrast, when it comes to either the harmonic or the Liouville-Weyl groupoids, the limit $d=4$ still sees the quadratic terms in $A_\mu$ coupling directly to the curvature tensors, even if $F_{\mu\nu}$ is introduced,
implying that invariance under the groupoids does not guarantee gauge $U(1)$ invariance in general. Such invariance should then be imposed separately for a gauge theory, if needed.
Therefore, in both $S_{\rm har} [A,g]$ and $S_{\rm LW} [A,g]$
the field $A_\mu$ is in general a Proca field, rather than a gauge field. It can be constrained further to be an Abelian gauge field,
but in that case the actions become fully Weyl invariant.

\subsection{Example: groupoids invariant hook-antisymmetric tensor actions}

Here we briefly present a first non-trivial example of the difference between the harmonic and Liouville subgroups when it comes to the dimension of the parameter spaces of the invariant actions. From the previous discussion, we deduce that such a difference only arises when we take into account mixed-symmetry tensors. We focus on a hook-antisymmetric traceless tensor $\kappa^\rho{}_{\mu\nu}$, i.e.,
\begin{equation}
 \kappa^\rho{}_{[\mu\nu]} = \kappa^\rho{}_{\mu\nu} \, , \qquad \kappa^\rho{}_{\mu\rho} = 0 \, , \qquad \kappa_{[\rho\mu\nu]} = 0 \, .
\end{equation}
The conformally invariant action of $\kappa^\rho{}_{\mu\nu}$ is known \cite{Paci:2023twc}, and it provides one of the simplest cases in which there are two independent conformally invariant contractions with the Weyl tensor.
The starting point is
\begin{align}\nonumber
S_0[g,\kappa] = \frac{1}{2} \int {\rm d}^dx \sqrt{g} & \Bigl\{ (\nabla_\alpha \kappa^\rho{}_{\mu\nu}) \nabla^\alpha \kappa_\rho{}^{\mu\nu} +\gamma_1 (\nabla_\alpha \kappa^\alpha{}_{\mu\nu}) \nabla_\beta \kappa^{\beta\mu\nu} +\gamma_2 (\nabla_\alpha \kappa^{\mu\alpha\nu}) \nabla_\beta \kappa_\mu{}^\beta{}_\nu  \Bigr\}\, ,
\end{align}
which is scale-invariant. The vector $j_\mu$ can be written as the divergence of a symmetric tensor $J_{\mu\nu}$ only if
$\gamma_1=- \frac{4(d-4)}{d^2-4}$ and $\gamma_2=- \frac{8}{d-2}$
as shown in Ref.~\cite{Paci:2023twc}.
The explicit form of the symmetric tensor then reads
\begin{equation}
 J^{\lambda\rho} = \frac{d-2}{2} g^{\rho\lambda} \kappa^2 + \kappa^\lambda{}_{\alpha\beta} \kappa^{\rho\alpha\beta} + 2 \kappa_\alpha{}^\lambda{}_\beta \kappa^{\alpha\rho\beta} \, . 
\end{equation}
We split it into the invariant components
\begin{eqnarray}
 X_0^{\rho\lambda} = \kappa^\rho{}_{\alpha\beta} \kappa^{\lambda\alpha\beta} + 2 \kappa^{\alpha\rho\beta} \kappa_\alpha{}^\lambda{}_\beta - \frac{3}{d} g^{\rho\lambda} \kappa^2 \, , && \qquad \Phi_0 = \frac{d^2-2d+12}{4d} \kappa^2 \, .
\end{eqnarray}
Therefore, the harmonic invariant quadratic action for the hook antisymmetric traceless tensor is
\begin{align}\nonumber
 S_{\rm har}[g,\kappa] = \frac{1}{2} \int {\rm d}^dx \sqrt{g} & \Bigl\{ \nabla_\alpha \kappa^\rho{}_{\mu\nu} \nabla^\alpha \kappa_\rho{}^{\mu\nu} - \frac{4(d-4)}{d^2-4} \nabla_\alpha \kappa^\alpha{}_{\mu\nu} \nabla_\beta \kappa^{\beta\mu\nu} - \frac{8}{d-2} \nabla_\alpha \kappa^{\mu\alpha\nu} \nabla_\beta \kappa_\mu{}^\beta{}_\nu \\\nonumber
 & + \Bigl( \kappa^\mu{}_{\alpha\beta} \kappa^{\nu\alpha\beta} + 2 \kappa^{\alpha\mu\beta} \kappa_\alpha{}^\nu{}_\beta - \frac{3}{d} g^{\mu\nu} \kappa^2 \Bigr) K_{\mu\nu} + \varsigma_1 \kappa^2 \mathcal{J}\\
 & + \rho_1 \kappa^\alpha{}_{\mu\nu} \kappa_{\alpha\rho\sigma} W^{\mu\nu\rho\sigma} + \rho_2 \kappa_{\mu\alpha\rho} \kappa_\nu{}^\alpha{}_\sigma W^{\mu\nu\rho\sigma} \Bigr\} \, ,
\end{align}
for arbitrary couplings $\rho_i$ and $\varsigma_i$.
The Liouville-Weyl invariant action is instead
\begin{align}\nonumber
S_{\rm LW}[g,\kappa] = \frac{1}{2} \int {\rm d}^dx \sqrt{g} & \Bigl\{ \nabla_\alpha \kappa^\rho{}_{\mu\nu} \nabla^\alpha \kappa_\rho{}^{\mu\nu} - \frac{4(d-4)}{d^2-4} \nabla_\alpha \kappa^\alpha{}_{\mu\nu} \nabla_\beta \kappa^{\beta\mu\nu} - \frac{8}{d-2} \nabla_\alpha \kappa^{\mu\alpha\nu} \nabla_\beta \kappa_\mu{}^\beta{}_\nu \\\nonumber
& + \varsigma_2 \Bigl( \kappa^\mu{}_{\alpha\beta} \kappa^{\nu\alpha\beta} - \frac{1}{d} g^{\mu\nu} \kappa^2 \Bigl) K_{\mu\nu} + \varsigma_3 \Bigl( \kappa^{\alpha\mu\beta} \kappa_\alpha{}^\nu{}_\beta - \frac{1}{d} g^{\mu\nu} \kappa^2 \Bigr) K_{\mu\nu}\\
& + \frac{d^2-2d+12}{4d} \kappa^2 \mathcal{J} + \rho_1 \kappa^\alpha{}_{\mu\nu} \kappa_{\alpha\rho\sigma} W^{\mu\nu\rho\sigma} + \rho_2 \kappa_{\mu\alpha\rho} \kappa_\nu{}^\alpha{}_\sigma W^{\mu\nu\rho\sigma} \Bigr\} \, ,
\end{align}
also for arbitrary couplings $\rho_i$ and $\varsigma_i$.
The kinematical ghosts, that are also present in the flat-space limit of the conformal action for the hook-antisymmetric traceless tensor \cite{Paci:2023twc}, cannot evidently be removed by relaxing the symmetry requirement to one of the subgroups. Indeed, this is a general feature of all harmonic or Liouville-Weyl invariant theories: in the flat-space limit their kinematical spectrum is the same as that of their conformal counterparts. This is because invariance under the two subgroups holds only when the coefficients of the kinetic terms in $S_0$ are the same as those found by requiring conformal invariance.

\section{Weyl groupoids as partial gauge fixings of full Weyl invariance}
\label{sect:partial-gf}

One interesting property of the harmonic groupoid of the Weyl group is that it can be seen as the invariance of a gauge-fixed actions which arise from a \emph{partial} gauge fixing of full conformal/Weyl invariance \cite{Oda:2020wdd,Edery:2023hxl}.
Partial gauge fixings are possible also in more traditional gauge theories and can be seen as related to the leftover gauge invariances that are present when gauge fixing with a covariant gauge such as Feynman's \cite{Oda:2020wdd}. In the case of the harmonic groupoid,
the partial gauge fixing is particularly interesting because it unveils some less-explored relations between models of higher derivative gravity, as we discuss in due time.

\subsection{From Weyl to harmonic Weyl: linear BRST}

In order to partially gauge fix a Weyl invariant action to a harmonic Weyl invariant action, we must choose a gauge fixing that is not invariant under the full Weyl group, but that instead is invariant only under the harmonic one. We concentrate here on the case $d=4$ because of its physically interesting applications,
though most of what we show can work in any $d$. The obvious choice for the gauge fixing is the curvature scalar $R$ itself, which has been discussed in Ref.~\cite{Oda:2020wdd}. In this section we give a linear realization of the BRST, while in the next section we improve it to a nonlinear form, which respects the Weyl weight of the fields involved in the construction. However, the final result of the two constructions is the same.

We can build an off-shell BRST transformation $\delta_B$ introducing a scalar ghost $c$, a scalar antighost $\overline{c}$ and a Nakanishi-Lautrup field $b$, such that
\begin{equation}\label{eq:harmonic-brst}
 \delta_B g_{\mu\nu} = 2 c g_{\mu\nu}\,,
 \qquad
 \delta_B c =0\,,
 \qquad
 \delta_B \overline{c} = b\,,
 \qquad
 \delta_B b=0\,,
\end{equation}
and it is straightforward to check that $\delta_B^2=0$ using the anticommuting nature of the ghosts (the field $b$ is instead commuting).
It is convenient to compute the transformations of $R$ and $\sqrt{g}$ before proceeding any further, that are
\begin{equation}\label{eq:harmonic-brst2}
 \delta_B R = -6 \Box_g c - 2 R c\,,
 \qquad
 \delta_B \sqrt{g} = 4 \sqrt{g} c\,,
\end{equation}
for which we have also used $d=4$ explicitly.

Having in mind that $R$ is the gauge-fixing function, we have that the combination of gauge fixing and ghost actions can be defined as the BRST transformation of the functional
\begin{equation}\label{eq:gf-gh-action-harmonic}
 S_{\rm gf}+S_{\rm gh}
 =
 \delta_B \int{\rm d}^4 x \sqrt{g} \,\overline{c}\,\Bigl(R-\frac{\alpha}{2}b \Bigr)
 =
 \delta_B \psi[g,b]
 \,,
\end{equation}
where $\alpha$ is a constant gauge-fixing parameter and the necessary functional $\psi[g,b]$ is implicitly defined.
Acting with $\delta_B$
we obtain, using Eqs.~\eqref{eq:harmonic-brst} and \eqref{eq:harmonic-brst2}, the following variation
\begin{equation}
\delta_B \psi[g,b]
 =
\int{\rm d}^4 x \sqrt{g} \Bigl( b  R - \frac{\alpha}{2} \, b^2- 2 \alpha  c \overline{c} b  + 6  \,\overline{c}\, \Box_g c  + 2 R \overline{c} c \Bigr)
 \, .
\end{equation}
Accordingly, the Nakanishi-Lautrup field has simple algebraic equations of motion
\begin{equation}
 b = \frac{1}{\alpha} R + 2 \overline{c}c \,,
\end{equation}
due to its nature as the auxiliary field, that allows to have BRST invariance even off-shell. If we use the equations of motion of $b$ inside \eqref{eq:gf-gh-action-harmonic}, the BRST remains valid, but only in its on-shell nonlinear form.
Doing so, we find
\begin{equation}\label{eq:gf-gh-action-harmonic-onshell}
  S_{\rm gf}+S_{\rm gh}
 =\frac{1}{2\alpha}\int{\rm d}^4 x \sqrt{g} R^2 + 6 \int{\rm d}^4 x \sqrt{g}
 \,\overline{c}\, \Box_g c 
 \,,
\end{equation}
which can be added to any fully Weyl invariant action to partially gauge fix Weyl symmetry to its harmonic subgroup.

The most important application of this construction concerns higher-derivative conformal gravity (HDCG). The action of HDCG is
\begin{equation}\label{eq:hdcg-action}
  S_{\rm wg}
 =\int{\rm d}^4 x \sqrt{g} \Bigl\{
 \frac{1}{2\lambda} W^2 + \frac{1}{2\eta} E_4
 \Bigr\}\,,
\end{equation}
where $W^2$ is the square of the Weyl tensor and $E_4$ is the four-dimensional Euler topological density \cite{deBerredo-Peixoto:2003jda}.
The action \eqref{eq:hdcg-action} is manifestly diffeomorphism and Weyl invariant up to boundary terms. Including the partial gauge fixing procedure
through the terms \eqref{eq:gf-gh-action-harmonic-onshell}, we have the gauge fixed version of $S_{\rm wg}$ 
\begin{equation}\label{eq:action-hdg-full}
  S = S_{\rm wg} + S_{\rm gf}+S_{\rm gh}
 =\int{\rm d}^4 x \sqrt{g} \Bigl\{
 \frac{1}{2\lambda} W^2
 + \frac{1}{2\eta} E_4
 + \frac{1}{2\alpha} R^2
 \Bigr\}
 + 6 \int{\rm d}^4 x \sqrt{g}\overline{c} \Box_g c
 \,.
\end{equation}
In the first three terms we recognize the standard form of the higher derivative gravity (HDG) action
\begin{equation}
  S_{\rm hdg}
 =\int{\rm d}^4 x \sqrt{g} \Bigl\{
 \frac{1}{2\lambda} W^2
 + \frac{1}{2\eta} E_4
 + \frac{1}{2\alpha} R^2
 \Bigr\}\,,
\end{equation}
which is the non-Weyl invariant version of $S_{\rm wg}$,
so that $S=S_{\rm hdg}+S_{\rm gh}$ and $S_{\rm hdg}=S_{\rm wg} + S_{\rm gf}$. The important point is that, in the partially gauge fixed action of HDCG, the constant $\alpha$ is interpreted as a gauge-fixing parameter, while, in the HDG action, $\alpha$ is generally interpreted a coupling. This difference in the interpretation of $\alpha$, and potentially of its (non)renormalization, is amended by the presence of the scalar ghosts. The action $S$, given in \eqref{eq:action-hdg-full},
is not fully Weyl invariant, but it is invariant under the harmonic Weyl group, so the discussions of Sect.~\ref{sect:intro} apply.

\subsection{A general nonlinear BRST for harmonic Weyl}

In an attempt to full generality, we now construct a consistent nonlinear version of the BRST given in \eqref{eq:harmonic-brst}. Its advantages appear in due time.
The most general action of $\delta_B$ on the fields should be of the form
\begin{equation}
 \delta_B g_{\mu\nu} = 2 c g_{\mu\nu}\,,
 \qquad
 \delta_B c =0\,,
 \qquad
 \delta_B \overline{c} =-2 c \overline{c} + \alpha_2 b +  \alpha_3 R\,,
 \qquad
 \delta_B b= -2 c b + \alpha_4 c R + \alpha_5  \Box c  \,,
\end{equation} 
which is a general ansatz with coefficients $\alpha_i$ that need to be fixed consistently.
We have already chosen $\delta_B c =0$, because it is fixed by the requirement $\delta_B^2 g_{\mu\nu} =0$, and the form of $\delta_B g_{\mu\nu}$ is fixed by Weyl transformations (generalized by replacing the infinitesimal conformal transformation $\sigma$ with the anticommuting ghost field $c$).

In the ansatz, there is more freedom when acting on the antighost and the $b$ fields, in fact $\delta_B \overline{c}$ includes now the term $-2c\overline{c}$ that takes into account the Weyl weight of $\overline{c}$,
which is naturally $w_{\overline{c}}=-2$. Similarly the weight of $b$ is accounted for in $\delta_B b$.
By imposing nilpotency of the BRST transformation (that is, $\delta_B^2=0$) on all the fields,
it is easy to show that the only consistent choices for three parameters are $\alpha_3=\alpha_4=\alpha_5=0$. Instead, the parameter $\alpha_2$ is not fixed, but can be changed at will by rescaling the field $b$, so we set $\alpha_2=1$ without loss of generality. The transformations simplify to
\begin{equation}
 \delta_B g_{\mu\nu} = 2 c g_{\mu\nu}\,,
 \qquad
 \delta_B \overline{c} =-2 c \overline{c} + b \,,
 \qquad
 \delta_B c =0 \,,
 \qquad
 \delta_B b= -2 c b \,,
\end{equation} 
and are consistent also with the weights of $\overline{c}$ and $b$. Importantly, in comparison to \eqref{eq:harmonic-brst}, the nonlinear version correctly attributes the weights
$w_{\overline{c}}=-2$ and $w_{b}=-2$
to the antighost and the auxiliary field.
Using the same definition as \eqref{eq:gf-gh-action-harmonic} for gauge-fixing 
and ghost actions, we obtain once more the algebraic equations of motion for the Nakanishi-Lautrup field, i.e.,
\begin{equation}
 b=\frac{1}{\alpha} R \,,
\end{equation} 
which are simpler in comparison to the previous section and are also consistent with the BRST transformations.
In this way, we ultimately find the same gauge fixing and ghost actions as \eqref{eq:gf-gh-action-harmonic-onshell}.

\subsection{The renormalization of conformal vs higher derivative gravity and the ghosts}

The renormalization of HDG with the action $S_{\rm hdg}$ is known to lead to an asymptotically free running for the coupling $\lambda$, but a trivial running for $\alpha$ when concentrating to the branch in which the low-energy limit gives a correct Newtonian potential \cite{Salvio:2018crh}.\footnote{The running of $\alpha$ is trivial in the sense that perturbatively it has a positive beta function as long as $\alpha>0$. Asymptotic freedom occurs for $\alpha<0$ instead \cite{deBerredo-Peixoto:2004cjk} and the fractal properties of the resulting quantum gravitational theory are heavily affected by the sign of $\alpha$ \cite{Becker:2019fhi}.} Recall that in that case $\alpha$ must be interpreted as a coupling. In all applications of HDG the triviality of $\alpha$ must always be circumvented.
On the other hand, the conformal counterpart HDCG, only has the asymptotically free running of $\lambda$ (and no coupling $\alpha$ to begin with). In both cases, $\eta$ is an asymptotically free coupling and the theories might suffer from unitarity issues because of the higher derivative propagators.

The relation among the action, i.e., $S=S_{\rm hdg}+S_{\rm gh}$,
can be seen at the level of the beta functions and it is very illuminating.
The beta function of $\lambda$ in HDCG is $\beta_\lambda = -\frac{1}{(4\pi)^2}\frac{199}{15}\lambda^2$, while in standard HDG it is $\beta_\lambda = -\frac{1}{(4\pi)^2}\frac{133}{10}\lambda^2$. It has already been noted in the past that the two results differ exactly by the contribution of two scalars \cite{Salvio:2017qkx}, which have been interpreted as the decoupled conformal modes at high energies when a running from HDG to HDCG is established. Such difference in the running is the essence of the agravity proposal of \cite{Salvio:2017qkx} and, from this point of view, would be interpreted as the renormalization group of HDG flowing towards \emph{higher} energies,
where the ultraviolet theory would be controlled by simply HDCG, which is asymptotically free.

However, as we know from Wilson's work \cite{Wilson:1983xri}, the renormalization group
only flows towards \emph{lower} energies, that is, only towards the infrared,
because it represents a loss of information on the system.\footnote{The renormalization group always contains a coarse-graining of the microscopic degrees of freedom, even though this might not be evident when computing the flow under the assumptions of dimensional regularization, which is valid only in-between thresholds \cite{Buccio:2023lzo}.} The construction with the partial gauge fixing hints at a different point of view to the agravity proposal: in the ultraviolet the dynamics is controlled by the asymptotically free HDCG (without having to decouple conformal modes). Then, HDG emerges as a partially gauge fixed version of HDCG, which is consistent to HDCG thanks to the presence of the scalar ghost and antighost.
From the point of view of the partial gauge fixing procedure, the running is always the same, at high energies we have asymptotic freedom (due to conformal gravity), while the $R^2$ term is just a gauge-fixing one whose contribution is precisely cancelled by the effect of two scalar ghosts.
The gauge-fixing parameter $\alpha$ should not renormalize, although this has not yet been tested with the model at hand.

\section{Generalizing paths to Weyl restrictions}
\label{sect:generalization}

A natural generalization of the discussion given in Sect.~\ref{sect:constructions} goes in the direction of constructing nonlinear homogeneous constraints on $\Omega$
of higher degree, i.e., with more than two derivatives. 
One may follow a similar approach and, starting from a linear combination of operators homogeneous in $\Omega$, look for constraints that are linear combinations satisfying the associative semigroupoids property, to obtain a set of restricting conditions at a given order. In this section, instead, we follow a different path, which automatically generates homogeneous conditions starting from the conformal properties of some tensors, and leads to proper associative composition properties for the resulting semigroupoids.

We first discuss the procedure in general terms, and then apply it to generate the family of scalar conditions with four derivatives, which can realize new restrictions to the Weyl group, and give also an interesting scalar constraint with six derivatives.

\subsection{Generating the basis of constraining differential operators}
\label{subsec6A}

We start from a linear space of tensors in a (pseudo)Riemannian manifold, as discussed in Section~\ref{sect:intro} and
focus on the general transformation properties of a Weyl noninvariant tensor under the Weyl rescalings $g_{\mu\nu} \to g'_{\mu\nu}= \Omega^2 g_{\mu\nu}$.
We adopt a compact notation to avoid displaying too many indices. To fix ideas, we have in mind the algebra of tensors constructed from curvatures, covariant derivatives and their contractions.
Notice that one could generalize the space of tensors introducing the dependence on other fields besides the metric (i.e., a Weyl connection that transforms affinely as, for example, in Ref.~\cite{Sauro:2022chz,Zanusso:2023vkn}, but also other non geometrical (matter) fields), but this goes beyond our current purposes.

For any given tensor $T$ that is not Weyl covariant, with a given scaling dimensions $\Delta$ (which is fixed by the number of derivatives acting on the metric), we can define a homogeneous
nonlinear ``shift'' operator $O_g[\Omega]$ as follows
\begin{equation}\label{rep_weyl}
T[g] \overset{\Omega}{\longrightarrow} T'[g]=T[g']=T[\Omega^2 g]=\Omega^{-\Delta} \left(  T[g] +O_g[\Omega]\right)\,,
\end{equation}
which constitutes a representation of the Weyl group on the space of the corresponding tensor. Notice that, using the above definition, $\Delta$ is essentially the negative of the Weyl weight of $T$ and coincides with the usual definition of scaling dimension of an operator for a scalar $T$.

If we consider the composition of two finite Weyl transformations associated to the conformal factors $\Omega_1$ and $\Omega_2$, then we can infer how the associativity of the Weyl group is realized in the given representation.
It is straightforward to derive the composition of two Weyl transformations as in Eq.~\eqref{rep_weyl}, first with factor $\Omega_1$ and then with factor $\Omega_2$, which gives
\begin{equation}\label{associative1}
T(\Omega_1^2g)  \overset{\Omega_2}{\longrightarrow}  T(\Omega_1^2 \Omega_2^2 g) =\Omega_2^{-\Delta} \left(  T[\Omega_1^2g] +O_{\Omega_1^2g}[\Omega_2]\right)
=\Omega_2^{-\Delta} \left(  \Omega_1^{-\Delta} \left(  T[g] +O_g[\Omega_1]\right) +O_{\Omega_1^2g}[\Omega_2]\right)\,.
\end{equation}
On the other hand, a Weyl transformation associated to the factor $\Omega_1\cdot\Omega_2$ is, by definition, given by
\begin{equation}\label{associative2}
T(\Omega_1^2 \Omega_2^2 g) =\left(\Omega_1\Omega_2\right)^{-\Delta} \left(  T[g] +O_g[\Omega_1 \Omega_2]\right)\,.
\end{equation}
These expressions must coincide by the associativity of the general Weyl group and, using also the fact that the Weyl group is Abelian, we have as a consequence that the following relations for the homogeneous nonlinear tensor operator $O_g$ are automatically satisfied
\begin{equation}\label{associative}
O_g[\Omega_1 \Omega_2] = O_g[\Omega_1] +\Omega_1^{\Delta}  O_{\Omega_1^2g}[\Omega_2] = O_g[\Omega_2] +\Omega_2^{\Delta}  O_{\Omega_2^2g}[\Omega_1] \,.
\end{equation}
We also have that $O_g[1]=0$, because $T$ must not transform for the trivial Weyl rescaling, so we can find the expression for the inverse transformation easily
\begin{equation}\label{inverse}
O_{ g}[\Omega^{-1}] = - \Omega^{-\Delta}  O_{\Omega^{-2} g}[\Omega] \,.
\end{equation}
We stress that all these definitions are related to a given metric $g_{\mu\nu}$.
In general one can write an infinite number of decompositions of a Weyl transformation, which induce the same amount of decompositions for the associated shift-operator $O_g[\Omega]$:
\begin{equation}\label{sequence}
O_g[\prod_{i=1}^n \Omega_i] = \sum_{i=1}^n \left( \prod_{j=1}^{i-1} \Omega_j^\Delta\right)  
O_{ \left( \prod_{j=1}^{i-1} \Omega_j^2\right) g}[\Omega_i]  \,.
\end{equation}

Clearly, linear combinations of tensors of the same rank and which transform according to the same scaling dimension $\Delta$ imply that the shift-operators $O_g(\Omega)$ can be linearly combined.
Therefore, a constraint realized by $O_{g}[\Omega]=0$ defines a Weyl substructure which leaves invariant the corresponding $T[g]$ up to a rescaling, and the same can be said for conditions and tensors constructed by
linear combinations, provided that they share the same scaling dimension $\Delta$ and rank. One can notice that, by associativity, each operator $O_{ \left( \prod_{j=1}^{i-1} \Omega_j^2\right) g}[\Omega_i]$ acting on $\Omega_i$ depends on the product of all the $\Omega_j$, for $j<i$.
Another useful piece of information comes from the transformation of the product of two tensors, which allows to study the full tensorial algebra. Considering a family of tensors $T_i[g]$,
each transforming under a Weyl rescaling of the metric with factor $\Omega$ according to Eq.~\eqref{rep_weyl} with scaling dimension $\Delta_i$ and shift-operators $O_{i,g}[\Omega]$,
one can write
\begin{equation}\label{product}
(T_iT_j)[\Omega^2 g] =\Omega^{-(\Delta_1+\Delta_2)} \Bigl(  (T_iT_j)[g] +O_{ij,g}[\Omega] \Bigr) \,,
\end{equation}
where
\begin{equation}\label{Opinproduct}
O_{ij,g}[\Omega] = T_i[g] O_{j ,g}[\Omega] +T_j[g] O_{i ,g}[\Omega] +O_{i ,g}[\Omega] O_{j ,g}[\Omega]\,,
\end{equation}
that defines a representation of the Weyl group.

Let us stress that, while the objects $O_{i ,g}[\Omega]$ and $O_{ij,g}[\Omega]$ satisfy the associative composition law in Eq.~\eqref{associative}, the product $O_{i ,g}[\Omega] O_{j ,g}[\Omega] \ne O_{ij,g}[\Omega]$ typically does not.
A case of particular interest is when $i=j$, which gives
\begin{equation}\label{Opsquare}
O_{ii,g}[\Omega] =2 T_i[g] O_{i ,g}[\Omega] +O_{i ,g}[\Omega]^2=O_{i ,g}[\Omega]\Bigl( O_{i ,g}[\Omega]+2T_i[g]\Bigr)  \,.
\end{equation}
Clearly these considerations can be extended to any product and power of tensors.

At this point we can consider the definition of possible restrictions of the Weyl group by imposing differential constraints of the form $O_g[\Omega]=0$, which transform correctly according to the associative law, and obtain a semigroupoid structure. 
One could, in principle, impose more than one constraint, if the related system of differential equations admits nontrivial solutions.

Choosing a starting metric $g_{\mu\nu}$, one can construct a submanifold of the space of metrics, which is associated to a the specific Weyl restricion, i.e., to the corresponding groupoid action. 
Taking a sequence of restricted Weyl transformations such as the ones represented in Eq.~\eqref{sequence} requires to impose 
$O_{ \left( \prod_{j=1}^{i-1} \Omega_j^2\right) g}[\Omega_i] = 0$ for any $i=1,2 \cdots n$, making it clear that each $\Omega_i$ solution
in the sequence depends on the product of the previous step solutions $\Omega_j$ for $j<i$.  The choice of the set of $\Omega_j$ defines a path in the restricted submanifold.
The simplest case corresponds to the constraint $O_g(\Omega)=0$ proportional to a linear operator, such as, for example, the case $O_g(\Omega) \propto \Box_g \Omega$ corresponding to the harmonic Weyl condition in $d=4$. Then one can easily construct, as already discussed in Section~\ref{sect:harmonic-group}, infinitesimal scale transformations, in the sense that the $\Omega_i$ are arbitrarily close to the identity $\Omega=1$, e.g., using perturbation theory. 
However, this can indeed be done also for a generic nonlinear constraints, at the price of dealing with more involved expressions.

\subsection{Scalar conditions with four derivatives}

We are interested here in the form of the possible differential scalar constraints which contains four derivatives and generalize the Harmonic-Weyl scalar constraint with two derivatives, which is related to the scalar Ricci curvature $R$ and scales with $\Delta=2$ as was discussed in Section~\ref{sect:intro}.
We stress that, in the following, our procedure will focus on deriving constraints obtained from variations, and ignore other possible cases which, for example, would generalize the case of the lightcone constraint for two derivatives, discussed in Sect.~\ref{lightcone_case}.

Our starting point is to consider all possible independent scalar quantities containing four derivatives of the metric. 
Taking into account that the Weyl tensor $W_{\mu\nu\alpha\beta}$ scales homogeneously under Weyl transformations using the definition
\begin{equation}
W_{\mu\nu\rho\sigma}W^{\mu\nu\rho\sigma}=
R_{\mu\nu\rho\sigma}R^{\mu\nu\rho\sigma}-\frac{4}{d-2} R_{\mu\nu}R^{\mu\nu}+\frac{2}{(d-1)(d-2)} R^2\,,
\end{equation}
one can deduce that, for this specific linear combination, there is no corresponding shift-operator ${}_4O_g$ of fourth order.
Therefore we can write only three linearly independent scalar operators with four derivatives studying the Weyl rescaling of $T_1=\Box R$, $T_2=R^2$ and $T'_3=R_{\mu\nu}R^{\mu\nu}$. Actually, we find more convenient to consider, instead of the Ricci tensor squared, the square of its traceless counterpart, that is, $T_3=\tilde{R}_{\mu\nu}\tilde{R}^{\mu\nu}=T'_3-\frac{1}{d} T_2$.

Under Weyl transformations, the three scalars $T_i$ for $i=1,2,3$ are characterized by the same scaling dimension $\Delta=4$.
They generate three shift-operators $ {}_4O_{i ,g}[\Omega]$ for  $i=1,2,3$, which are a basis of a three dimensional space of $4$-derivative constraints. 

One or more of these constraints may be used to define Weyl restricted semigroupoids.
Using the relations given above we can write explicitely the three operators
\begin{eqnarray}\label{opsca4der}
\Omega^4 O_{1,g}[\Omega]&=& -2(d-1)\Omega^3 \Box_g^2 \Omega+
2\Omega \Box_g\Omega\Bigl( 3(d-1) \Omega \Box_g\Omega -R\Omega^2 +
(d-1)(5d-26) \partial_{\alpha }\Omega \partial^{\alpha }\Omega \Bigr) \nonumber \\
&{}&-2(d-1)(d-4) \Bigl[ \Omega \nabla_{\mu } \partial_{\nu }\Omega 
\Bigl( \Omega \nabla^{\mu } \partial^{\nu }\Omega+(d-10) \partial^{\mu }\Omega \partial^{\nu }\Omega
\Bigr)+\Omega^2 R_{\mu\nu}\partial^{\mu }\Omega \partial^{\nu }\Omega \Bigr] \nonumber \\
&{}&+(d-6)\Omega^3 \partial_{\mu }R  \partial^{\mu }\Omega-2(d-5) R \Omega^2 \partial_{\mu }\Omega \partial^{\mu }\Omega 
+ 4(d-1)(d-4)(d-7) (\partial_{\mu }\Omega \partial^{\mu } \Omega )^2
\nonumber \\
&{}&-4(d-1)(d-6) \Omega^2 \partial_{\mu } \Omega \partial^{\mu }\Box_g \Omega \,, 
\label{opsca4der1}\\ \nonumber \\
\frac{\Omega^4}{d-1}  O_{2,g}[\Omega]&=&
4 \Omega H_g(\Omega)\Bigl[ (d-1) \Omega H_g(\Omega)-R\, \Omega^2\Bigr]  
\label{opsca4der2}\\
 &=&
\Bigl(2 \Omega \Box_g\Omega + (d-4)  \partial_{\alpha }\Omega \partial^{\alpha }\Omega\Bigr) 
\Bigl[ (d-1)\Bigl(2 \Omega \Box_g\Omega + (d-4)  \partial_{\beta}\Omega \partial^{\beta }\Omega\Bigr)-2R \Omega^2 \Bigr]\,,
\nonumber \\ \nonumber \\
\frac{\Omega^4}{d-2}  O_{3,g}[\Omega]&=&
\Omega^5 L_{g,\mu \nu}(\Omega)\Bigl[ (d-2) \Omega L_{g,\alpha \beta}(\Omega)-2 \tilde{R}_{\alpha \beta}\Bigr] g^{\mu \alpha}g^{\nu \beta} \label{opsca4der3}\\ 
&=& \frac{(d-2)}{d} \Omega \Box_g\Omega  \Bigl[ (-\Omega \Box_g\Omega +4 \partial_{\alpha }\Omega \partial^{\alpha }\Omega \Bigr] +\frac{2}{d}  \Omega^3 R \,\Box_g\Omega  \nonumber
\\&{}&
+\frac{4}{d}\partial_{\alpha }\Omega \partial^{\alpha }\Omega  \Bigl[(d-1)(d-2) \partial_{\beta }\Omega \partial^{\beta }\Omega 
- \Omega^2 R  \Bigr] \nonumber
-2 \Omega^2 R_{\mu\nu} \Bigl(\Omega \nabla^{\mu } \partial^{\nu }\Omega-2 \partial^{\mu }\Omega \partial^{\nu }\Omega \Bigr)
\nonumber \\&{}&
+(d-2) \Omega \nabla_{\mu } \partial_{\nu }\Omega
\Bigl(\Omega \nabla^{\mu } \partial^{\nu }\Omega-4 \partial^{\mu }\Omega \partial^{\nu }\Omega \Bigr) \nonumber \,.
\end{eqnarray}
Recall that $\partial^\mu = g^{\mu\nu}\partial_\nu$ and $\nabla_\mu\Omega = \partial_\mu \Omega$.

The second condition $O_{2,g}[\Omega]$ has a convenient factorized form into the product of two two-derivative terms, where one of the factors is precisely the scalar harmonic condition. This comes from considering the shift-operator associated to the Ricci scalar $R$, given by $O_{R,g}[\Omega]=-2\frac{d-1}{\Omega} H_g(\Omega)$, after using Eq.~\eqref{Opsquare}. Similarly, the third condition can be derived immediately from the form of the shift-operator associated to the traceless Ricci tensor $\tilde{R}_{\mu\nu}$, given by 
$O_{\tilde{R}_{\mu\nu},g}[\Omega]=-(d-2)\Omega L_{g,\mu\nu}(\Omega)$.
The first condition, instead, has a more involved and apparently irreducible expression.
In general, one can write a general differential scalar constraint in the form
\begin{equation}
{}_4O_g[\Omega]=\sum_{i=1}^3 c_i O_{i,g}[\Omega]\,.
\end{equation}
Imposing ${}_4O_g[\Omega]=0$, one can ignore also an overall factor, so that the differential constraint depends on two free real parameters.
Solutions of this equation may or may not exist, depending on these parameters and on $g_{\mu\nu}$, which therefore could characterize the semigroupoid structure.
In particular, keeping as nonzero the coefficient of $O_{1,g}[\Omega]$, which contains the highest derivative $\Box_g^2\Omega$ term, one can define
a two parameter family of differential constraints  
${}_4 H_{g; a, b}[\Omega]=O_{1,g}[\Omega]+a \,O_{2,g}[\Omega]+b \,O_{3,g}[\Omega]$.

One interesting constraint is obtained from the analysis of the Weyl rescaling of the scalar $Q$-curvature $Q_4$, imposing the condition that it does not change apart from a simple rescaling.
These curvatures are motivated by the study of conformal geometry
and, roughly, are defined as tensors with simple conformal transformations \cite{Gover:2002ay}.
For general $d$, a simple expression for this scalar is given by~\cite{Osborn:2015rna}
\begin{equation}
Q_4=\frac{d}{2}{\cal J}^2-2K^{\mu\nu}K_{\mu\nu}-\nabla^2{\cal J}\,, 
\end{equation}
Here $K^{\mu\nu}$ is the Schouten tensor defined in Eq.~\eqref{eq:schouten-def} and ${\cal J}=\frac{1}{2(d-1)} R$ its trace.
Rewriting $Q_4$ in terms of the scalar basis given above together with the Weyl square scalar, $\{T_1, T_2, T_3, W^2\} $, 
\begin{equation}
Q_4=-\frac{1}{2(d-1)} \Box R+\frac{(d-2)(d+2)}{8d(d-1)^2} R^2-\frac{2}{(d-2)^2} \tilde{R}_{\mu\nu}\tilde{R}^{\mu\nu} \,,
\end{equation}
one immediately obtains the related scalar constraint in terms of the ones given in Eq.~\eqref{opsca4der1}, \eqref{opsca4der2} and \eqref{opsca4der3}, as
\begin{equation}\label{eq:q4-Constraint}
O_{Q_4,g}[\Omega]=-\frac{1}{2(d-1)} O_{1,g}[\Omega]+
\frac{(d-2)(d+2)}{8d(d-1)^2} O_{2,g}[\Omega]-\frac{2}{(d-2)^2} O_{3,g}[\Omega]\,.
\end{equation}
The form of $O_{Q_4,g}$ acquires a remarkably simple form in $d=6$
\begin{equation}\label{eq:q4-Constraint-6d}
O_{Q_4,g}[\Omega]=\frac{1}{\Omega} \left[ \Box_g^2 +R_{\mu\nu} \nabla^{\mu } \nabla^{\nu }-\frac{1}{2} R\, \Box_g \right] \Omega\,.
\end{equation}
This is a linear condition in $\Omega$ (since the overall $\Omega^{-1}$ is factored out), in complete analogy with the harmonic scalar constraint $H_g$, proportional to what obtained varying $Q_2= {\cal J} $, which becomes simpler and linear in $d=4$.

\subsubsection{Example: higher derivatives theories invariant under  the $Q_4$-restriction.}

Here
we shall briefly consider two simple examples of scalar theories that are invariant under the constraint given in Eq.~\eqref{eq:q4-Constraint}, with four and six derivatives actions.

The first example, in general dimensions $d\ge4$, is obtained from a deformation of a Weyl-invariant quadratic action contructed with the Paneitz-Riegert-Fradkin-Tseytlin operator $\Delta_4$
\begin{align}
S[\varphi, g]=\frac{1}{2}\int {\rm d}^d x\sqrt{g}\,\varphi\left(\Delta_4+\xi Q_4\right)\varphi\,,
\end{align}
where $\varphi$ is a scalar field with weight $w_\varphi=\frac{4-d}{2}$ and the Weyl-covariant Paneitz operator reads \cite{paneitz}
\begin{align}
\Delta_4 = \square_g^2 + \nabla_\mu\left[4K^{\mu\nu}-(d-2)g^{\mu\nu}{\cal J}\right]\nabla_\nu + \frac{d-4}{2}Q_4\,.
\end{align}
For $\xi=0$, the above action is Weyl invariant by construction, while, for an arbitrary $\xi$, the infinitesimal variation of the action is proportional to the linearized version of Eq.~\eqref{eq:q4-Constraint}
\begin{align}
O_{{\rm lin},g}[\sigma] = \left[\square_g^2-\frac{d^2-4d+8}{2(d-2)(d-1)}R\,\square_g +\frac{4}{d-2}R^{\mu\nu}\nabla_\mu\nabla_\nu
		-\frac{d-6}{2(d-1)}\nabla^\mu R\nabla_\mu \right]\sigma\,.
\end{align}
Since the linearized constraint is self-adjoint up to boundary terms in generic dimensions, i.e., $O_{{\rm lin},g}=O^{\dagger}_{{\rm lin},g}$, it is easy to explicitly check that the trace of the energy-momentum tensor has the following form up to total derivatives
\begin{align}
 T^\mu_\mu = \frac{\xi}{2}O_{{\rm lin},g}[\varphi^2]
 +\frac{d-4}{2}\varphi\left(\Delta_4+\xi Q_4\right)\varphi\,,
\end{align}
where the last term on the right vanishes on-shell.

We have checked the form of the $1$-loop anomaly in $d=6$ (which is cubic in the curvatures) using the result of Ref.~\cite{Casarin:2023ifl}, expecting a shift from the well-known anomaly of $\Delta_4$ proportional to $\xi\, R\, Q_4$. However, due to a cancellation between the terms in the expression of the heat kernel coefficient $b_6(\Delta_4)$, the trace anomaly remains of the same form as the well-known anomaly for a scalar propagating with $\Delta_4$.\footnote{In particular, for a higher derivative operator of the form $\Delta=\square^2 + V^{\mu\nu}\nabla_\mu\nabla_\nu+2N^\mu\nabla_\mu+U$, a shift $\delta U$ of the endomorphism in $d=6$ amounts to a variation in the heat kernel coefficient of the form $\delta b_6(\Delta)={\rm tr}\left[\left(-\frac{1}{12}V^\mu_\mu-\frac{1}{6}R\right)\delta U\right]$. See also Ref.~\cite{Casarin:2023ifl}.}

It is interesting to notice that, in $d=6$, we can construct a shift-invariant action from a higher derivative operator of rank six, that is symmetric under the constraint \eqref{eq:q4-Constraint-6d}. To this end, let us consider, as second example, a scalar field $\varphi$ with zero weight under conformal transformation with dynamics governed by the following action
\begin{align}\label{eq:shift-higher-dvs}
 S[\varphi, g]=\frac{1}{2}\int {\rm d}^6x\sqrt{g}\, \left(\varphi\Delta_6\varphi
 +\xi \,g^{\mu\nu}\partial_\mu\varphi\partial_\nu\varphi\,Q_4\right)\,.
\end{align}
Here, $\Delta_6$ is the GJSM operator constructed out of six derivatives which becomes shift invariant in six dimensions since the $Q_6$-curvature is absent in $d=6$ (see, e.g., the appendix of Ref.~\cite{Osborn:2015rna} and references therein). As before, the action is Weyl invariant when $\xi$ vanishes and the rescaling of $Q_4$ is now compensated by the measure and the inverse metric. The peculiarity of the above action is that the energy-momentum tensor has the expected form without the need of going on-shell, i.e., the following property holds off-shell up to boundary terms
\begin{align}
T^\mu{}_\mu = \frac{\xi}{2} \, O_{lin, g}[\partial_\mu\varphi \partial^\mu\varphi]\,.
\end{align}
However, the computation of the anomaly for such a case is much more complicated, since it is related to the rank six differential operator $\Delta_6$, also deformed by $Q_4 (\partial \varphi)^2$. Covariant computational tools for high rank operators are much less known (as compared to rank the two and four cases), so we postpone its analysis to future works.

We stress that while, according to the previous section, there exist infinitely many scalar restrictions with four derivatives,
one also expects potentially interesting cases related to restrictions obtained from two-indices tensors that we have not investigated here. These would include generalizations of the Liouville-Weyl grupoid, which may play a useful role in (shift-symmetric) theories in six and higher dimensions.

\subsection{An example of scalar condition with six derivatives}

In this subsection, we show an example of a six derivatives constraint obtained by considering the Weyl transformation of the $Q$-curvature $Q_6$. In general, there exist $17$ linearly independent purely geometrical scalars with $6$ derivatives acting on the metric, among which $3$ can be chosen to be Weyl invariant combinations and $7$ are boundary terms (i.e., total derivatives) \cite{Bastianelli:2000rs}. For the sake of readability, we omit many details of the calculations that are not particularly illuminating and focus on the result that we want to present. In general spacetime dimension $d$, this scalar can be expressed using conformal tensors as~\cite{Osborn:2015rna}
\begin{equation}
\begin{split}
 Q_6  =&\,
 \frac{16}{d-4} B_{\mu\nu} K^{\mu\nu} + \frac{(d-2)(d+2)}{4}  {\cal J} ^3
 -4d {\cal J}  K_{\mu\nu} K^{\mu\nu}
 +16 K^{\mu\nu}K_{\mu\sigma} K_{\nu}{}^\sigma
 \\&
 -\frac{d+10}{2}  {\cal J} \Box {\cal J} 
 +8 K^{\mu\nu}\nabla_\mu\nabla_\nu  {\cal J} 
 + \Box \Big(4 K_{\mu\nu} K^{\mu\nu} -  \frac{d-6}{2} {\cal J} ^2\Big)
  +\Box^2  {\cal J}  \, ,
\end{split}
\end{equation}
where we also notice the obstruction in $d=4$ for nonvanishing Bach tensor $B_{\mu\nu}$ \cite{Gover:2002ay}. 
We follow the definitions of \cite{Osborn:2015rna}, so in our notation the Bach tensor is defined as
\begin{equation}
B_{\mu\nu}=\nabla^\rho C_{\mu\nu\rho}-K^{\rho \sigma}W_{\rho\mu\nu\sigma}\,,
\end{equation}
where $C_{\mu\nu\rho}$ is the Cotton tensor, which is written in terms of the Schouten tensor as
\begin{equation}
C_{\mu\nu\rho}= \nabla_\rho K_{\mu\nu}- \nabla_\nu K_{\mu\rho}\,.
\end{equation}
The above expression for $Q_6$ can be expanded as a linear combination of $10$ linearly independent scalars.

Proceeding in a complete analogy to the previous section, it is possible compute in general dimension the associated scalar constraint $O_{Q_6,g}$, whose expression is quite involved and we do not write it for brevity. Nevertheless, the expression is greatly simplified in $d=8$, becoming proportional to a linear operator acting on $\Omega$, which takes the form
\begin{equation}
\setlength{\jot}{8pt}
\begin{split}
O_{Q_6,g}[\Omega]
&=
-\frac{1}{\Omega} \bigg[
 \Box^3 
+ 16   \tilde{K}^{\mu\nu} \nabla_\mu\nabla_\nu \Box
- 7 {\cal J} \Box^2 
+ 16  (\nabla^\alpha \tilde{K}^{\mu\nu} ) \nabla_\alpha  \nabla_\mu     \nabla_\nu 
- 12   {\cal J}   \tilde{K}^{\mu\nu} \nabla_\mu\nabla_\nu \\
&  
+\frac{63}{4} {\cal J}^2  \Box
-\frac{7}{2} (\Box {\cal J}) \Box
+144  \tilde{K}_{\alpha}{}^{\mu} \tilde{K}^{\alpha\nu}  \nabla_\mu\nabla_\nu 
- 8 \tilde{K}_{\mu\nu} W^{\mu\alpha\nu\beta}  \nabla_\alpha \nabla_\beta  
- 28  \tilde{K}^{\mu\nu}  \tilde{K}_{\mu\nu} \Box   \\
&
+7 {\cal J}  (\nabla^\alpha {\cal J} ) \nabla_\alpha
+12 ( \Box  \tilde{K}^{\mu\nu} ) \nabla_\mu\nabla_\nu 
+ 56   \tilde{K}^{\mu\nu} (\nabla_\mu {\cal J}) \nabla_\nu
+ 128 \tilde{K}_{\mu\nu}  (\nabla^\nu \tilde{K}^{\alpha\mu} ) \nabla_\alpha \\
&
- 64 \tilde{K}^{\mu\nu}  (\nabla^\alpha \tilde{K}_{\mu\nu} ) \nabla_\alpha \bigg]
\Omega
\, ,
\end{split}
\end{equation}
where $\tilde{K}_{\mu\nu}=K_{\mu\nu} -\frac{g_{\mu\nu}}{d} {\cal J} $ is the traceless Schouten tensor. The limit $d=8$ of the constraint coming from $Q_6$ gives a linear operator in complete analogy to the $d=6$ case coming from $Q_4$, and also to the harmonic case in $d=4$. We thus conjecture that the shift-operator coming from the transformation of the $2n$-th $Q$-curvature generates a linear differential constraint in $d=2n+2$, but we have no proof for this statement yet.

\section{Conclusions}
\label{sect:conclusions}

We have provided a classification of some group-like substructures of the Weyl group that are based on differential constraints on the conformal factor that rescales the metric. We believe that our classification of the two-derivative constraints is complete under a reasonable set of assumptions and certainly covers all the interesting cases that are of use for four dimensional theories.
The main actors that emerge from our analysis are the harmonic and the Liouville-Weyl subgroups of the Weyl group, though, as we explained in the main text, they are actually partial associative groupoids, since each transformations depends on the metric itself.

The substructures of the Weyl group are interesting because, in terms of the properties that they imply, they are somewhere in between scale and conformal invariance, displaying more free couplings than fully Weyl-invariant theories. For example, the trace of the energy-momentum tensor of a theory invariant under harmonic or Liouville-Weyl transformations in nonzero, but it is still constrained to be the Laplacian of a scalar or the double divergence of a traceless tensor, respectively. This means that any invariant model can always be improved to have a fully traceless energy-momentum tensor, however
this comes at the price of losing some freedom in the available couplings.
Reasonably, the flat space limit of theories invariant under the subgroups gives conformal theories just like full Weyl invariance does, although this should be confirmed by further explorations beyond the simple analysis of the quantum anomaly, which we have carried out only in few simple examples.

The most interesting context in which the substructures can be applied is, in our opinion, that of conformal gravity. In fact, we can perform a partial gauge fixing of Weyl symmetry of higher derivative conformal gravity to show that, in the case of the harmonic Weyl subgroup, the model can be recast as the ``sum'' of the action of standard higher derivative gravity and that of two scalar ghosts. Higher derivative gravity is known to be asymptotically free in the coupling with the Weyl tensor, which is fully Weyl invariant, but not in the coupling with the curvature scalar squared, at least for values that give a reasonable Newtonian potential. The partial gauge fixing allows one to see the latter coupling as a gauge-fixing term, which is thus not renormalized, as long as one is willing to include the two scalar ghosts. In this way, it becomes natural to think at higher derivative conformal gravity, which is always asymptotically free, as a natural completion of gravity in the ultraviolet, while the standard higher derivative gravity becomes its partially gauge-fixed counterpart. Of course, the analysis that we have presented does not solve the issue of nonunitarity of higher derivative models, for which one should adopt one of the several recommended solutions, e.g., Refs.~\cite{Mannheim:2020ryw,Mannheim:2021oat,Anselmi:2018ibi}, but, maybe, it can be helpful in further developing the program of completing theories of gravity as ultraviolet consistent theories, especially now that physical UV complete trajectories of the renormalization group have been found in higher derivative (quadratic) gravity \cite{Buccio:2024hys}.

We have also found a third, nontrivial, potentially interesting two-derivative constraint of lightcone type, which leads to a new unexplored Weyl subgroup, but we have not yet found interesting application to models enjoying such symmetry. Null hypersurfaces are known to have codimension two
and are generally used to discuss the physics of black holes, for example, but are also relevant in the context of asymptotic symmetries.

Finally, we have presented a way to generalize the construction of Weyl-restricted subgroups by enforcing differential constraints with an arbitrary number of derivatives and have given some results for four-derivative scalar constraints, which are typically very non linear. Among them, the specific constraint that is obtained by Weyl variations of a $Q$-curvature, denoted $Q_4$ \cite{Gover:2002ay,Osborn:2016bev}, becomes linear in $d=6$ and takes a very simple expression. To explore if this is a generalizable feature, we have also constructed a six-derivative scalar constraint from the variation of $Q_6$, which also becomes linear in $d=8$, which leads to our conjecture that this is a general property: varying $Q_{2n}$ gives a linear operator in $d=2n+2$ which can act as a definition of a groupoid substructure of the Weyl group. Although the higher-dimensional structures are more exotic, it could be that they become relevant in the context of conformal geometry \cite{Gover:2002ay}.

\appendix

\section{Partial associative groupoids}\label{sect:appendix-groupoids}

Let ${\cal W}$ be the group of Weyl transformations, i.e., the set of nonzero functions $\Omega(x)$ that map $g_{\mu\nu}\to g'_{\mu\nu}=\Omega^2 g_{\mu\nu}$. As a set we have
$$
{\cal W} = \bigcup_{\Omega} \bigl\{g_{\mu\nu} \to \Omega^2 g_{\mu\nu}; ~{\rm such ~that}~ \Omega >0  \bigr\}
$$
and $\Omega \in {\cal F}$ the space of functions over the spacetime manifold.
An element $f\in {\cal W} $ induces the Weyl transformation
$$f: g_{\mu\nu} \to \Omega^2 g_{\mu\nu}$$
for some $\Omega$ on \emph{any} metric that it acts on.
The group multiplication is induced by the pointwise multiplication of the functions, so if $a: g_{\mu\nu} \to \Omega_1^2 g_{\mu\nu}$ and $b: g_{\mu\nu} \to \Omega_2^2 g_{\mu\nu}$, then $ a \circ b:g_{\mu\nu} \to \Omega_2^2\cdot \Omega_1^2 g_{\mu\nu}$. The action inherits the Abelian nature of the product of functions, $a\circ b =b \circ a$, and its associativity, $a\circ (b\circ c)=(a\circ b)\circ c$, for any $a,b,c\in {\cal W}$.
The identity element is $e: g_{\mu\nu} \to g_{\mu\nu}$ and we always have an inverse, e.g., $f^{-1}: g_{\mu\nu} \to \Omega^{-2} g_{\mu\nu}$ such that $f\circ f^{-1}= f^{-1}\circ f =e$.

The harmonic Weyl group ${\cal W}_R$ is a groupoid given by the set of maps such that
$$
{\cal W}_R = \bigcup_{g,\Omega} \bigl\{g_{\mu\nu}\to \Omega^2 g_{\mu\nu}; ~{\rm such ~that}~ \Omega >0 ~{\rm and}~ H_{g}(\Omega)=0  \bigr\}
$$
with $H_g(\Omega)$ defined in Eq.~\eqref{eq:harmonic}.
Notice that when singling out elements of ${\cal W}_R$ we have to clarify which metric the conformal transformation acts on, for example $g_{\mu\nu}$, besides the conformal factor $\Omega$, so ${\cal W}_R$ is not a subset of ${\cal W}$, but rather of the union of all $g_{\mu\nu}$s and all $\Omega$s.

Two elements of ${\cal W}_R$ can be composed if and only if the second attaches to the image metric of the first, making the composition operation \emph{partial}
because it does not apply to an arbitrary pair of the \emph{total} set.\footnote{The operation does not satisfy the \emph{totality} axiom.}
For example if $a,b\in {\cal W}_R$ for $a=\bigl\{g_{\mu\nu}\to \Omega_1^2 g_{\mu\nu} | H_{g}(\Omega_1)=0  \bigr\}$ and $b=\bigl\{\Omega_1^2 g_{\mu\nu}\to \Omega_2^2 \cdot \Omega_1^2 g_{\mu\nu} | H_{\Omega_1^2 g}(\Omega_2)=0  \bigr\}$, then the composition gives a third element $ b \circ a$ such that $b \circ a=\bigl\{g_{\mu\nu}\to \Omega_2^2\cdot \Omega_1^2 g_{\mu\nu} | H_{g}(\Omega_2\cdot \Omega_1)=0\bigl\}$.
Since there is an internal product, the set ${\cal W}_R$ is a groupoid (a.k.a., a magma), but it is only a \emph{partial groupoid} because the product does not extend to any two elements of ${\cal W}_R$.

The properties of $H_g(\Omega)$ under composition ensures associativity, i.e., $a\circ (b\circ c)=(a\circ b)\circ c$, for any $a,b,c\in {\cal W}_R$ that can be composed in the above order, making ${\cal W}_R$ a partial associative groupoid, i.e., what is often referred to as a \emph{partial semigroupoid}.\footnote{The general definition of partial semigroupoid is somewhat vague in the math literature, because
it often depends on the specific applications. So take this statement as our working definition. See Refs.~\cite{wikipedia,mathoverflow} and references within.}
The groupoid does not have an identity, but rather it has identity \emph{elements} $e_g=\bigl\{g_{\mu\nu}\to g_{\mu\nu}\bigr\}\in {\cal W}_R$ because, strictly speaking, there is one identity element for each metric (the set union and identification of all these identity elements would then give the identity of the full Weyl group). For any given element $f=\bigl\{g_{\mu\nu}\to g'_{\mu\nu}=\Omega^2 g_{\mu\nu} | H_{g}(\Omega)=0  \bigr\} \in {\cal W}_R$, the groupoid also has the ``right-inverse'' map
$f^{-1}=\bigl\{g'_{\mu\nu}=\Omega^2 g_{\mu\nu}\to  g_{\mu\nu} | H_{\Omega^2 g}(\Omega^{-1})=0  \bigr\}\in {\cal W}_R$, such that $f^{-1}\circ f =e_g$. The right-inverse $f^{-1}$ is guaranteed to exist because the differential equations $H_{g}(\Omega)=0$ and $H_{\Omega^2 g}(\Omega^{-1})=0$ are completely equivalent as long as $\Omega>1$, so if $f$ exists also $f^{-1}$ does. However, $f^{-1}$ is not the proper inverse of a group, as discussed in the main text. For example not only $f\circ f^{-1}\neq e_{g'}$, but it generally does not even make sense to compose $f$ and $f^{-1}$ in that order as $f$ might not attach properly to the image of $f^{-1}$ because the groupoid is partial.
It remains true, however, that $f\circ e_g = e_{g'} \circ f =f$, although it should be noted that the identity elements on the two sides of the equations act on different metrics.

An exactly equivalent discussion holds for the other groupoids of the Weyl group,
in particular the Liouville-Weyl introduced in Sect.~\ref{sect:intro} 
and the ones associated to the higher order differential conditions discussed in Sect.~\ref{sect:generalization}, all of
which enjoy as defining properties closure and associativity, making them partial semigroupoids.

\section{Cohomology of finite transformations}\label{sect:appendix-cohomology}

One way to further validate the results of the analysis of the quantum anomaly that have been obtained in Sect.~\ref{subsubsect:anomaly-example-1} would be to inspect the Wess-Zumino consistency conditions \cite{Osborn:1991gm}. However, such analysis cannot be carried out straightforwardly in the case of the harmonic Weyl group (and, in general, of any partial semigrupoid). The problem is rooted in the differential constraint imposed on the bosonic parameter function $\Omega$, which forbids the application of two transformations in the ``opposite'' order, as required to construct a commutator. To better visualize what goes wrong consider the following diagram:
\begin{center}
\begin{tikzpicture}[commutative diagrams/every diagram]
\node (P0) at (90:2.3cm) {$g_{\mu\nu}$};
\node (P1) at (90+72:2.5cm) {\small{$g_1=\Omega^2_1 g, \, H_g(\Omega_1)=0$}} ;
\node (P2) at (55+2*72:2.5cm) {\makebox[26ex][r]{\small{$g_{12}=\Omega^2_2 g_1, \, H_{g_1}(\Omega_2)=0 $ }}};
\node (P3) at (125+3*72:2.5cm) {\makebox[20ex][l]{\small{$g_{21}=\Omega^2_1 g_2, \,  H_{g_2}(\Omega_1)\neq0$}}};
\node (P4) at (90+4*72:2.5cm) {\small{$g_2=\Omega^2_2 g, \, H_g(\Omega_2)\neq0$}};
\path[commutative diagrams/.cd, every arrow, every label]
(P0) edge node[swap] {$\Omega_1$} (P1)
(P1) edge node[swap] {$\Omega_2$} (P2)
(P4) edge node {$\Omega_1$} (P3)
(P0) edge node {$\Omega_2$} (P4);
\end{tikzpicture}
\end{center}
\vspace{0.5cm}
Moving to the left, we need to require $H_g(\Omega_1)=0$ and $H_{g_1}(\Omega_2)=0 $, which conflicts with the conditions on the $\Omega_i$s that we should impose if instead moving to the right, i.e., $ H_g(\Omega_2)=0$ and $H_{g_2}(\Omega_1)=0$.
Therefore, $\Omega_{1,2}$ on the right cannot be the same as those on the left
if we want to restrict ourselves to the harmonic groupoid. Furthermore, another difficulty is that we should resort to a \emph{finite} version of the transformations of the harmonic Weyl group, as done in \cite{Mazur:2001aa}, rather than to an infinitesimal one, due to the lack of an algebra.

One possible way out is to perform a cohomological analysis, in which we regard  $\sigma=\log \Omega$ as Grassmannian anticommuting function, and proceed in a way similar to what we have done for the BRST transformations in Sect.~\ref{sect:partial-gf}
with a nilpotent transformation $\delta_\sigma$.
For bosonic $\sigma$ we cannot follow the right path of the diagram for the same reasons explained above, but the finite transformation is now possible since $\Omega=1+\sigma$, by virtue of the anticommuting nature of $\sigma$.
To frame the discussion, consider the simple example of a weight $w_\varphi=-1$ scalar that is nonminimally coupled to the metric in $d=4$. We start by defining the operator $\delta_\sigma$ as in \cite{Bonora:1985cq}
\begin{equation}
\delta_\sigma
=
\delta^{g}_\sigma
+
\delta^{\varphi}_\sigma
=
2\int {\rm d}^4x \, \sigma \, g_{\mu\nu} \frac{\delta}{\delta g_{\mu\nu}}
+ w_\varphi
\int {\rm d}^4x \, \sigma\,  \varphi \frac{\delta}{\delta \varphi}\,,
\end{equation}
which is a proper co-boundary operator, i.e., it is nilpotent $\delta^2_\sigma=0$, when $\sigma$ is an anticummuting function. This operator is nilpotent like the exterior derivative in differential geometry, so that we can construct its cohomology. Following Ref.~\cite{Bonora:1985cq}, the consistency conditions are then
\begin{equation}\label{eq:1-cocycle}
\delta_\sigma \omega[\sigma, g]
=
0 \, ,
\end{equation}
where $\omega$ can be written as a linear combination of all possible one-cochains that are built using derivatives, metric and inverse metric, and that are linear in $\sigma$. Then, the condition \eqref{eq:1-cocycle} of $\omega$ being closed imposes constraints on the coefficients that appear in $\omega$ itself, which ultimately select the admissible tensor structures of the anomaly.

Among these structures, some are cohomologically trivial in the sense that they automatically satisfy \eqref{eq:1-cocycle}, i.e., they are exact one-cochains. Therefore, the actualy nontrivial anomalies are those that belong to the analog of the De Rham cohomology, built using $\delta_\sigma$ instead of the exterior differential. However, since we are only interested in justifying the structures appearing in Sect.~\ref{subsubsect:anomaly-example-1}, we leave aside the general task of classifying the cohomology group, and concentrate only on the consistency conditions. To do so, we start by constructing a convenient basis in the case of harmonic Weyl symmetry for the possible one-cochains of dimension four 
\begin{align}\label{eq:1-cochains}
&\omega_1 =\int {\rm d}^4x \sqrt{g} \sigma W^{\alpha\beta\rho\gamma} W_{\alpha\beta\rho\gamma} \, ,
&&\omega_5 =\int {\rm d}^4x \sqrt{g} \sigma \varphi^4\,,
\nonumber\\
&\omega_2 =\int {\rm d}^4x \sqrt{g} \sigma E_4  \, ,
&&\omega_6 =\int {\rm d}^4x \sqrt{g} \sigma R \varphi^2\,,
\nonumber\\
&\omega_3 =\int {\rm d}^4x \sqrt{g} \sigma R^2   \, ,
&&\omega_7=\int {\rm d}^4x \sqrt{g} \sigma \Box \varphi^2\,,
\nonumber\\
&\omega_4 = \int {\rm d}^4x \sqrt{g} \sigma   \Box R \, ,
&&\omega_8=\int {\rm d}^4x \sqrt{g} \sigma   \varphi  \Box \varphi \, ,
\nonumber
\end{align}
where $E_4$ is the four dimensional Euler density. Therefore, we can parametrize $\omega=c_i \omega_i[\sigma, g]$ and thus Eq.~\eqref{eq:1-cocycle} becomes
\begin{equation}
\delta_\sigma \sum_{i} c_i \omega_i[\sigma, g] 
=
0 \, .
\end{equation}

In the main text the anomaly is found on-shell and at fixed points of the ronormalization group, so we can focus on the purely geometric contributions $\omega_i$, $i=1,\dots,4$. By computing the variation $\delta_\sigma$ of these terms, and using the anticommuting nature of $\sigma$ along with integration by parts, it is easy to see that the only nonzero contribution comes from $\omega_3[\sigma, g]$ and is given by
\begin{equation}
\delta_\sigma \omega_3[\sigma, g]
=
4 \int {\rm d}^4x \sqrt{g} R \sigma \Box \sigma \, .
\end{equation}
This result dictates that curvature squared terms must appear in the nontrivial anomaly only as $W^2$ and $E_4$ for the case of full Weyl group \cite{Bonora:1985cq}. In contrast, the harmonic Weyl condition relaxes this requirement, because, for Grassmannian but \emph{finite} $\sigma$, the harmonic condition becomes 
\begin{equation}
H_g(1+\sigma)=\Box_g \sigma=0
\, ,
\end{equation}
which allows for the presence of $R^2$ in the restricted Weyl anomaly as we saw from an explicitic computation in Eq.~\eqref{eq:weyl-restricted-anomaly-simple-scalar}. In this way its presence is also justified by the cohomological analysis.

In the case of the Liouville-Weyl anomaly it is more convenient to choose a different basis for the one-cochains, in which we replace $\omega_3[\sigma, g]$ with 
\begin{equation}
\omega'_3
=
\int {\rm d}^4x \sqrt{g} \sigma \tilde{R}_{\mu\nu}^2
\, .
\end{equation}
In this way we get 
\begin{equation}\label{eq:omega3-Liouville}
\delta_\sigma \omega'_3[\sigma, g]
=
4 \int {\rm d}^4x \sqrt{g}  \tilde{R}^{\mu\nu} \sigma \nabla_\mu \partial_\nu \sigma \, .
\end{equation}
Then, for $\sigma$ a finite Grassmanian function, the Liouville-Weyl constraints becomes
\begin{equation}
 L_{g,\mu\nu}(1+\sigma)
=
\Big( \nabla_\mu \partial_\nu -\frac{g_{\mu\nu}}{d}\Box_g \Big) \sigma
-
\partial_\mu \sigma \partial_\nu \sigma
=0
\, ,
\end{equation}
from which we see that \eqref{eq:omega3-Liouville} automatically vanish. Therefore, the presence of $ \tilde{R}_{\mu\nu}^2$ is cohomologically allowed, as expected.


\end{document}